%% file: rhombic_sampling_ius.tex
\documentclass[conference]{IEEEtran}

\usepackage{cite}
\usepackage{amsmath,amssymb,amsfonts}
\usepackage{algorithmic}
\usepackage{graphicx}
\usepackage{textcomp}
\usepackage{xcolor}
\def\BibTeX{{\rm B\kern-.05em{\sc i\kern-.025em b}\kern-.08em
    T\kern-.1667em\lower.7ex\hbox{E}\kern-.125emX}}

\usepackage[T1]{fontenc}
\usepackage[utf8]{inputenc}

\usepackage{scalerel}
\usepackage{flushend}

\usepackage{siunitx}
\usepackage{tikz}
\usepackage{pgfplots}
\pgfplotsset{compat=newest}
\usetikzlibrary{calc,intersections,svg.path,pgfplots.external,arrows.meta,backgrounds}
\usepgfplotslibrary{statistics}

\usepackage{array}	
\usepackage{booktabs}	
\usepackage{dcolumn}	
\usepackage{multirow}
\usepackage{multicol}

\usepackage{marginnote}
\newcommand{\TODO}[1]{}

\newcolumntype{H}{>{\scriptsize}c}

\usepackage[labelformat=simple,labelfont=bf,font=footnotesize,singlelinecheck=true]{subcaption}	

\DeclareCaptionLabelSeparator{periodspace}{.\quad}
\definecolor{orcidlogocol}{HTML}{A6CE39}
\tikzset{
  orcidlogo/.pic={
    \fill[orcidlogocol] svg{M256,128c0,70.7-57.3,128-128,128C57.3,256,0,198.7,0,128C0,57.3,57.3,0,128,0C198.7,0,256,57.3,256,128z};
    \fill[white] svg{M86.3,186.2H70.9V79.1h15.4v48.4V186.2z}
                 svg{M108.9,79.1h41.6c39.6,0,57,28.3,57,53.6c0,27.5-21.5,53.6-56.8,53.6h-41.8V79.1z M124.3,172.4h24.5c34.9,0,42.9-26.5,42.9-39.7c0-21.5-13.7-39.7-43.7-39.7h-23.7V172.4z}
                 svg{M88.7,56.8c0,5.5-4.5,10.1-10.1,10.1c-5.6,0-10.1-4.6-10.1-10.1c0-5.6,4.5-10.1,10.1-10.1C84.2,46.7,88.7,51.3,88.7,56.8z};
  }
}

\newcommand\orcidlink[1]{\href{https://orcid.org/#1}{\mbox{\scalerel*{
\begin{tikzpicture}[yscale=-1,transform shape]
\pic{orcidlogo};
\end{tikzpicture}
}{|}}}}

\usepackage{hyperref}

\hypersetup{%
  pdfauthor={M. F. Schiffner},%
  pdftitle={Rhombic Grids Reduce the Number of Voxels in Fast Pulse-Echo Ultrasound Imaging},%
  pdfsubject={Lehrstuhl für Medizintechnik, Ruhr-Universität Bochum},%
  pdfkeywords={%
    rhombic grids, %
    optimal orthogonal grids, %
    bivariate sampling theorem, %
    fast ultrasound imaging, %
    voxel-based beamforming, %
    point scatterers, %
    spectral properties
  },%
  colorlinks=false,%
  hidelinks,
  linktocpage=true,%
  bookmarksopenlevel=1%
}

\usepackage{breakurl}   

\usepackage{acronym}    

\usepackage{cleveref}	  			

\crefname{chapter}{Chapter}{Chapters}		
\Crefname{chapter}{Chapter}{Chapters}		
\crefname{appendix}{Appendix}{Appendices}
\Crefname{appendix}{Appendix}{Appendices}
\crefname{section}{Sect.}{Sections}
\Crefname{section}{Section}{Sections}
\crefname{subsection}{Sect.}{Sections}
\Crefname{subsection}{Section}{Sections}
\crefname{subsubsection}{Sect.}{Sections}
\Crefname{subsubsection}{Section}{Sections}

\crefname{figure}{Fig.}{Figs.}
\Crefname{figure}{Figure}{Figures}

\crefname{table}{Table}{Tables}
\Crefname{table}{Table}{Tables}

\newcommand{\name}[1]{#1}

\definecolor{RUBBLUE}{cmyk}{1.000,0.500,0.000,0.600}	
\definecolor{RUBGREEN}{cmyk}{0.500,0.000,1.000,0.000}	
\definecolor{RUBGRAY}{cmyk}{0.030,0.030,0.030,0.100}	

\definecolor{RUBBLUE_RGB}{rgb}{0.000,0.208,0.377}
\definecolor{RUBGREEN_RGB}{rgb}{0.553,0.682,0.063}
\definecolor{RUBGRAY_RGB}{rgb}{0.906,0.906,0.906}
\definecolor{RUBGRAYDARK_RGB}{rgb}{0.588,0.588,0.588}

\definecolor{gray}{gray}{0.5}
\definecolor{gray70}{gray}{0.3}
\definecolor{gray60}{gray}{0.4}
\definecolor{gray40}{gray}{0.6}
\definecolor{gray30}{gray}{0.7}
\definecolor{gray20}{gray}{0.8}
\definecolor{gray10}{gray}{0.9}
\definecolor{gray05}{gray}{0.95}

\newcommand{\setsymbol}[1]{\ensuremath{\mathbb{#1}}}
\newcommand{\N}{\setsymbol{N}}

\newcommand{\Z}{\setsymbol{Z}}

\newcommand{\R}{\setsymbol{R}}

\newcommand{\dsetcons}[2]%
{%
\ensuremath%
 {%
  \ifthenelse{\equal{#2}{0}}{[ #1 ]}{}%
  \ifthenelse{\equal{#2}{1}}{\bigl[ #1 \bigr]}{}%
  \ifthenelse{\equal{#2}{2}}{\Bigl[ #1 \Bigr]}{}%
  \ifthenelse{\equal{#2}{3}}{\biggl[ #1 \biggr]}{}%
  \ifthenelse{\equal{#2}{4}}{\Biggl[ #1 \Biggr]}{}%
 }%
}

\newcommand{\dsetconsnonneg}[2]%
{%
\ensuremath%
 {%
  \ifthenelse{\equal{#2}{0}}{[ #1 ]_{0}}{}%
  \ifthenelse{\equal{#2}{1}}{\bigl[ #1 \bigr]_{0}}{}%
  \ifthenelse{\equal{#2}{2}}{\Bigl[ #1 \Bigr]_{0}}{}%
  \ifthenelse{\equal{#2}{3}}{\biggl[ #1 \biggr]_{0}}{}%
  \ifthenelse{\equal{#2}{4}}{\Biggl[ #1 \Biggr]_{0}}{}%
 }%
}

\newcommand{\vect}[1]{\ensuremath{\mathbf{#1}}}

\newcommand{\uvectcompsymbol}{\ensuremath{e}}

\newcommand{\uvectsymbol}{\ensuremath{\vect{\uvectcompsymbol}}}
\newcommand{\uvect}[1]{\ensuremath{\uvectsymbol_{#1}}}

\newcommand{\transsymbol}{\ensuremath{\mathrm{T}}}
\newcommand{\trans}[1]{\ensuremath{#1^{\transsymbol}}}

\newcommand{\diagsymbol}{\ensuremath{\mathrm{diag}}}

\newcommand{\ddiag}[2]%
{%
 \ensuremath
 {%
  \ifthenelse{\equal{#2}{0}}{\diagsymbol\{ #1 \}}{}%
  \ifthenelse{\equal{#2}{1}}{\diagsymbol\bigl\{ #1 \bigr\}}{}%
  \ifthenelse{\equal{#2}{2}}{\diagsymbol\Bigl\{ #1 \Bigr\}}{}%
  \ifthenelse{\equal{#2}{3}}{\diagsymbol\biggl\{ #1\biggr \}}{}%
  \ifthenelse{\equal{#2}{4}}{\diagsymbol\Biggl\{ #1\Biggr \}}{}%
 }%
}

\newcommand{\dabs}[2]
{%
 \ensuremath%
 {
  \ifthenelse{\equal{#2}{0}}{\lvert#1\rvert}{}%
  \ifthenelse{\equal{#2}{1}}{\bigl\lvert#1\bigr\rvert}{}%
  \ifthenelse{\equal{#2}{2}}{\Bigl\lvert#1\Bigr\rvert}{}%
  \ifthenelse{\equal{#2}{3}}{\biggl\lvert#1\biggr\rvert}{}%
  \ifthenelse{\equal{#2}{4}}{\Biggl\lvert#1\Biggr\rvert}{}%
 }%
}

\newcommand{\inprod}[2]{\ensuremath{\left\langle #1, #2 \right\rangle}}

\newcommand{\dinprod}[3]{%
\ensuremath%
 {%
  \ifthenelse{\equal{#3}{0}}{\langle #1, #2 \rangle}{}%
  \ifthenelse{\equal{#3}{1}}{\bigl\langle #1, #2 \bigr\rangle}{}%
  \ifthenelse{\equal{#3}{2}}{\Bigl\langle #1, #2 \Bigr\rangle}{}%
  \ifthenelse{\equal{#3}{3}}{\biggl\langle #1, #2 \biggr\rangle}{}%
  \ifthenelse{\equal{#3}{4}}{\Biggl\langle #1, #2 \Biggr\rangle}{}%
 }%
}
\newcommand{\dinprodr}[3]{%
\ensuremath%
 {%
  \ifthenelse{\equal{#3}{0}}{\langle #1, #2 \rangle_{\text{r}}}{}%
  \ifthenelse{\equal{#3}{1}}{\bigl\langle #1, #2 \bigr\rangle_{\text{r}}}{}%
  \ifthenelse{\equal{#3}{2}}{\Bigl\langle #1, #2 \Bigr\rangle_{\text{r}}}{}%
  \ifthenelse{\equal{#3}{3}}{\biggl\langle #1, #2 \biggr\rangle_{\text{r}}}{}%
  \ifthenelse{\equal{#3}{4}}{\Biggl\langle #1, #2 \Biggr\rangle_{\text{r}}}{}%
 }%
}

\newcommand{\kroneckerdeltasymbol}{\ensuremath{\delta}}
\newcommand{\kroneckerdelta}[2]{\ensuremath{\kroneckerdeltasymbol}_{#1, #2}}

\input{0_acronyms.tex}
\acresetall

\begin{document}

\pgfmathsetmacro{\NVoxelsOrthogonalUsual}{262144}
\pgfmathsetmacro{\NVoxelsOrthogonalOptimal}{57117}
\pgfmathsetmacro{\NVoxelsRhombic}{48703}

\pgfmathsetmacro{\NVoxelsRelDiffUsual}{ 81.4 }
\pgfmathsetmacro{\NVoxelsRelDiffOptimal}{ 14.7 }

\pgfmathsetmacro{\MeanSSIMIndexOptimal}{ 96.9 }
\pgfmathsetmacro{\MeanSSIMIndexRhombic}{ 96.6 }

\pgfmathsetmacro{\relRMSEOptimal}{ 6.4 }
\pgfmathsetmacro{\relRMSERhombic}{ 6.8 }

\input{copyright.tex}

\title{Rhombic Grids Reduce the Number of Voxels in Fast Pulse-Echo Ultrasound Imaging}

\author{%
  \IEEEauthorblockN{Martin F. Schiffner\,\orcidlink{0000-0002-4896-2757}}%
  \IEEEauthorblockA{%
    \textit{Chair of Medical Engineering}\\
    \textit{Ruhr-University Bochum}\\
    Bochum, Germany\\
    martin.schiffner@rub.de%
  }
}

\maketitle

\begin{abstract}
  \input{abstract.tex}
\end{abstract}
\acresetall

\begin{IEEEkeywords}
  rhombic grids,
  optimal orthogonal grids,
  bivariate sampling theorem,
  fast ultrasound imaging,
  voxel-based beamforming,
  point scatterers,
  spectral properties
\end{IEEEkeywords}

\section{Introduction}
\input{introduction/introduction.tex}

\section{Theory}
\input{theory/theory.tex}

\section{Methods}
\input{methods/methods.tex}

\section{Results}
\input{results/results.tex}

\section{Conclusion}
\input{conclusion/conclusion.tex}

\bibliographystyle{IEEEtran}

\end{document}

%% file: 0_acronyms.tex
\acrodef{AAPM}{American Association of Physicists in Medicine}		
\acrodef{ACB}{asyn\-chro\-nous compressed beamformer}			
\acrodef{ACA}{adaptive cross approximation}				
\acrodef{ACR}{American College of Radiology}				
\acrodef{ADC}{analog-to-digital conversion}				
\acrodef{AIC}{analog-to-information conversion}				
\acrodef{BIBO}{bounded-input, bounded-output}				
\acrodef{BIM}{\name{Born} iterative method}				
\acrodef{BP}{basis pursuit}						
\acrodef{BPDN}{basis pursuit denoise}					

\acrodef{BVP}{boundary value problem}					
\acrodef{CAR}{Canadian Association of Radiologists}			
\acrodef{CDF}{cumulative distribution function}				
\acrodef{CGP}{conjugate gradient pursuit}				
\acrodef{CNR}{contrast-to-noise ratio}					
\acrodef{CoSaMP}{compressive sampling matching pursuit}			
\acrodef{CPU}{central processing unit}					
\acrodef{CPV}{\name{Cauchy} principal value}				
\acrodef{CPWC}{coherent plane-wave compounding}				
\acrodef{CS}{compressed sensing}					

\acrodef{CSI}{contrast source inversion}				
\acrodef{CT}{x-ray computed tomography}					
\acrodef{CTTE}[CT-TE]{continuous-time ternary encoding}			
\acrodef{CW}{cylindrical wave}						
\acrodef{DAC}{digital-to-analog conversion}				
\acrodef{DAIC}{diagnostically acceptable irreversible compression}	
\acrodef{DAS}{delay-and-sum}						
\acrodef{DBIM}{distorted \name{Born} iterative method}			
\acrodef{DCT}{discrete cosine transform}				
\acrodef{DICOM}{Digital Imaging and Communications in Medicine}		

\acrodef{DFT}{discrete \name{Fourier} transform}			
\acrodef{DSFT}{discrete space \name{Fourier} transform}			
\acrodef{DWT}{discrete wavelet transform}				
\acrodef{DNA}{deoxyribonucleic acid}					
\acrodef{DRG}{German \name{Röntgen} Society}				
\acrodef{DRIM}{distorted \name{Rytov} iterative method}			
\acrodef{DRS}{digital random sampler}					
\acrodef{ERM}{exploding reflector model}				
\acrodef{ESR}{European Society of Radiology}				
\acrodef{EU}{European Union}						

\acrodef{FBP}{filtered backpropagation}					
\acrodef{FDA}{US Food and Drug Administration}				
\acrodef{FDBF}{frequency domain beamforming}				
\acrodef{FDT}{\name{Fourier} diffraction theorem}			
\acrodef{FEHM}{full extent at half maximum}				
\acrodefplural{FEHM}{full extents at half maximum}			
\acrodef{FFT}{fast \name{Fourier} transform}				
\acrodef{FIR}{finite impulse response}					
\acrodef{FLOP}{floating point operation}				
\acrodef{FMM}{fast multipole method}					
\acrodef{FOCUS}{fast object-oriented C++ ultrasound simulator}		

\acrodef{FOV}{field of view}						
\acrodefplural{FOV}{fields of view}					
\acrodef{FPDE}{fractional partial differential equation}		
\acrodef{FPS}{frames per second}					
\acrodef{FRI}{finite rate of innovation}				
\acrodef{FWHM}{full width at half maximum}				
\acrodefplural{FWHM}{full widths at half maximum}			
\acrodef{FWT}{fast wavelet transform}					
\acrodef{GCNR}[gCNR]{generalized contrast-to-noise ratio}		
\acrodef{GPU}{graphics processing unit}					
\acrodef{GWN}{Gaussian white noise}					
\acrodef{HSV}{hue, saturation, value}					

\acrodef{HTP}{hard thresholding pursuit}				
\acrodef{IHT}{iterative hard thresholding}				
\acrodef{IID}[i.i.d.]{independent and identically distributed}		
\acrodef{IQ}{in-phase and quadrature}					
\acrodef{IRLS}{iteratively reweighted least squares}			
\acrodef{ISP}{inverse scattering problem}				
\acrodef{ISR}{integrated sidelobe ratio}				
\acrodef{JPEG}{Joint Photographic Experts Group}			
\acrodef{JSM}{joint sparsity model}					
\acrodef{KK}{\name{Kramers}-\name{Kronig}}                              

\acrodef{LASSO}{least absolute shrinkage and selection operator}	
\acrodef{LDT}{\name{Laplace} diffraction theorem}			
\acrodef{LS}{\name{Lippmann}-\name{Schwinger}}				
\acrodef{LSB}{least significant bit}					
\acrodef{LTI}{linear time-shift invariant}				
\acrodef{MAP}{maximum \emph{a posteriori}}				
\acrodef{MCMC}{\name{Markov} chain Monte Carlo}				
\acrodef{MIRF}{material impulse response function}			
\acrodef{MLFMA}{multilevel fast multipole algorithm}			
\acrodef{MMV}{multiple measurement vectors}				

\acrodef{MOM}[MoM]{method of moments}					
\acrodef{MP}{matching pursuit}						
\acrodef{MP3}{MPEG Audio Layer III}					
\acrodef{MRI}{magnetic resonance imaging}				
\acrodef{MTF}{material transfer function}				
\acrodef{MV}{minimum variance}						
\acrodef{MVDR}{minimum-variance distortionless response}		
\acrodef{MWC}{modulated wideband converter}				
\acrodef{NESTA}{\name{Nesterov}'s algorithm}				
\acrodef{NSP}{null space property}					

\acrodef{NNZC}{number of nonzero components}				
\acrodef{OMP}{orthogonal matching pursuit}				
\acrodef{PAM}{pulse-amplitude modulation}				
\acrodef{PAT}{photoacoustic tomography}					
\acrodef{PC}{personal computer}						
\acrodef{PDE}{partial differential equation}				
\acrodef{PDF}{probability density function}				
\acrodef{PET}{positron emission tomography}				
\acrodef{PICMUS}{Plane-Wave Imaging Challenge in Medical Ultrasound}	
\acrodef{PSF}{point spread function}					

\acrodef{PSL}{peak sidelobe level}					
\acrodef{PSNR}{peak signal-to-noise ratio}				
\acrodef{PW}{plane wave}						
\acrodef{QCW}{quasi-cylindrical wave}					
\acrodef{QPW}{quasi-plane wave}						
\acrodef{QSW}{quasi-spherical wave}					
\acrodef{QQ}{quantile-quantile}						
\acrodef{RA}{randomly-apodized}						
\acrodef{RAD}{randomly-apodized and randomly-delayed}			
\acrodef{RADAR}[RADAR]{radio detection and ranging}			

\acrodef{RAM}{random-access memory}					
\acrodef{RCR}{Royal College of Radiologists}				
\acrodef{RD}{randomly-delayed}						
\acrodef{RF}{radio frequency}						
\acrodef{RIC}{restricted isometry constant}				
\acrodef{RIP}{restricted isometry property}				
\acrodef{RMSE}{root mean-squared error}					
\acrodef{ROI}{region of interest}					
\acrodefplural{ROI}{regions of interest}				
\acrodef{ROMP}{regularized \ac{OMP}}					
\acrodef{RS}{\name{Rayleigh}-\name{Sommerfeld}}                         

\acrodef{SaS}[S$\alpha$S]{symmetric $\alpha$-stable}			
\acrodef{SA}{synthetic aperture}					
\acrodef{SBL}{sparse Bayesian learning}					
\acrodef{SDK}{software development kit}					
\acrodef{SDR}{software-defined radio}					
\acrodef{SIIM}{Society for Imaging Informatics in Medicine}		
\acrodef{SISO}{single-input, single-output}				
\acrodef{SINR}{signal-to-interference-plus-noise ratio}			
\acrodef{SMI}{sample matrix inversion}					
\acrodef{SNR}{signal-to-noise ratio}					

\acrodef{SOMP}{simultaneous orthogonal matching pursuit}		
\acrodef{SoS}{sum of sincs}						
\acrodef{SP}{subspace pursuit}						
\acrodef{spark}{sparse rank}						
\acrodef{SPECT}{single photon emission computed tomography}		
\acrodef{SPG}{spectral projected gradient}				
\acrodef{SPGL1}[SPG$\ell_{1}$]{spectral projected gradient for $\ell_{1}$-minimization}	
\acrodef{SPLS}{split-projection least squares}				
\acrodef{SR}{sparse recovery}						
\acrodef{SRC}{\name{Sommerfeld} radiation condition}			

\acrodef{SSI}{supersonic shear imaging}					
\acrodef{SSIM}{structural similarity}					
\acrodef{STFT}{short-time \name{Fourier} transform}			
\acrodef{StOMP}{stagewise \ac{OMP}}					
\acrodef{StWGP}{stagewise weak \ac{CGP}}				
\acrodef{SVD}{singular value decomposition}				
\acrodef{TDMA}{time-division multiple access}				
\acrodef{TMSBL}[T-MSBL]{temporally correlated \ac{MMV}-based \ac{SBL}}  
\acrodef{TOF}{time-of-flight}						
\acrodefplural{TOF}{times-of-flight}					
\acrodef{TGC}{time gain compensation}					

\acrodef{THI}{tissue harmonic imaging}					
\acrodef{TPSF}{transform point spread function}				
\acrodef{TR}{\name{Tikhonov} regularization}				
\acrodef{TV}{total variation}						
\acrodef{UCA}{ultrasound contrast agents}				
\acrodef{UDT}{ultrasound diffraction tomography}			
\acrodef{UI}{ultrasound imaging}					
\acrodef{UoS}{union of subspaces}					
\acrodef{USA}{United States of America}					
\acrodef{UWB}{ultra-wideband}						

\acrodef{Xampling}{CS-Sampling}						

%% file: copyright.tex
\begin{titlepage}
\thispagestyle{empty}%
\noindent
{\huge
Rhombic Grids Reduce the Number of Voxels in\\[0.8ex]
Fast Pulse-Echo Ultrasound Imaging
}
\par
\vspace{36pt}
\noindent
{\large Martin F. Schiffner\par\vspace{12pt}
\noindent \href{http://www.mt.rub.de}{Chair of Medical Engineering}, Ruhr-University Bochum, 44801 Bochum, Germany}
\vspace{36pt}
\par
\noindent
{\bf Copyright notice:}\par\vspace{12pt}
\noindent
\copyright~2022~IEEE.
Personal use of
this material is
permitted.
Permission from
IEEE must be obtained for
all other uses, in
any current or
future media, including reprinting/republishing
this material for
advertising or
promotional purposes, creating
new collective works, for
resale or
redistribution to
servers or
lists, or
reuse of
any copyrighted component of
this work in
other works.
\par
\vspace{12pt}
\noindent
{\bf Full citation:}\par\vspace{12pt}
\noindent
2022 IEEE Int. Ultrasonics Symp. (IUS), Venice, Italy, Oct. 2022, pp. 1--4.
\par
\noindent
DOI: \href{https://doi.org/10.1109/IUS54386.2022.9958278}{10.1109/IUS54386.2022.9958278}
\par
\vspace{12pt}
\noindent
\href{https://ieeexplore.ieee.org/document/9958278}{Click here for IEEE Xplore}
\clearpage
\end{titlepage}

%% file: abstract.tex
Ultrafast imaging modes, such as
\ac{CPWC}, capture
a large \acl{FOV} in
a single pulse-echo measurement using
parallel receive focusing.
The number of
foci or, equivalently,
the number of
volume elements (voxels) in
the image determines
the computational costs and
the memory consumption of
the image formation.
Herein,
\SI{120}{\degree} rhombic grids are proposed to specify
the voxel positions and reduce
the number of
voxels in comparison to
orthogonal grids.
The proposed grids derive from
the bivariate sampling theorem and
the spectral properties of
the images formed by
the \acl{DAS} algorithm in
\ac{CPWC}.
A phantom experiment validated
the proposed grids and showed reductions in
the number of
voxels by
\SI{\NVoxelsRelDiffUsual}{\percent} and
\SI{\NVoxelsRelDiffOptimal}{\percent} in comparison to
the usual and
optimal orthogonal grids,
respectively.
Mean \acl{SSIM} indices above
\SI{\MeanSSIMIndexRhombic}{\percent} and
relative \aclp{RMSE} below
\SI{\relRMSERhombic}{\percent} confirmed
the visual equivalence of
all images after
interpolations to
the usual orthogonal grid.

%% file: introduction/introduction.tex
Ultrafast imaging modes, such as
\ac{CPWC}
\cite{article:MontaldoITUFFC2009}, use
software to form
complete images from
single pulse-echo measurements
\cite{article:TanterITUFFC2014}.
Image formation algorithms, such as
\ac{DAS}
\cite{article:PerrotUlt2021} or
constrained $\ell_{q}$-minimization
\cite{article:SchiffnerArxiv2019,article:BerthonPMB2018}, discretize
the object to be imaged using
point scatterers on
an orthogonal grid.
Each point scatterer represents
a volume element (voxel) of
the object.
The grid of
point scatterers exactly represents
the entire object if
the grid spacings are
sufficiently small.
The limited bandwidth of
the pulse-echo measurements, then, makes
the grid of
point scatterers indistinguishable from
the object.
The algorithms, however, must compute
the strengths of
all point scatterers.
The \ac{DAS} algorithm, for example, measures
the amplitude of
the echo received from
each scatterer by
parallel focusing.
Since
the number of
point scatterers determines
the computational costs and
the memory consumption,
it is important to find
grids that not only enable
an exact object representation but also minimize
the number of
voxels.

Herein,
rhombic grids are proposed to reduce
the number of
voxels and, thus,
the computational costs and
the memory consumption of
the image formation.
The spectral properties of
the images formed by
the \ac{DAS} algorithm in
\ac{CPWC} will be reviewed first.
These properties will subsequently be inserted into
the bivariate sampling theorem
\cite{article:PetersenInfCtrl1962} to derive
two grids that enable
an exact object representation.
The first grid is
an optimal orthogonal grid.
The second grid is
the proposed rhombic grid.
A phantom experiment will eventually show
the advantages of
the rhombic grid over
the usual and
optimal orthogonal grids.


%% file: theory/theory.tex
%
\begin{figure}[t!]
 \centering%
  \input{theory/figures/latex/theory_grid.tex}
 \caption{}
 \label{fig:V}
\end{figure}
C
{
 Discretization of
 the object function using
 point scatterers on
 a regular grid.
 The object function describes
 spatial fluctuations in
 the acoustic object properties.
 The pulse-echo measurements, owing to
 the limited bandwidth, cannot distinguish
 the grid of
 point scatterers from
 the object.
 %
}%
{theory_grid}

This paper exclusively treats
pulse-echo measurements with
steered \acp{PW}, as shown in
\cref{fig:theory_grid}.
A linear array with
the pitch $p$ serves as
transducer.
All grid types are
regular
(i.e.,
the shapes of
all voxels are
equal).

\subsection{Spectral Properties of Ultrasound Images}
\label{subsec:theory_spectral_properties}
\input{theory/theory_spectral_properties.tex}

\subsection{Sampling of Ultrasound Images}
\label{subsec:theory_sampling}
\input{theory/theory_sampling.tex}

%% file: theory/figures/latex/theory_grid.tex
%
\begin{tikzpicture}%
[
  font = \footnotesize,
  axes_r/.style = { ->, thick },
  element/.style = { draw = black, thin },
  FOV/.style = { draw = black!20, draw opacity = 1, thick, text opacity = 1, rounded corners = 2 },
  cyst/.style = {circle, inner sep = 0pt, draw = black, draw opacity = 1, thin, text opacity = 1, fill = gray20, fill opacity = 1},
  hyper/.style = {circle, inner sep = 0pt, draw = black, draw opacity = 1, thin, text opacity = 1},
  wire/.style = {circle, inner sep = 0pt, minimum size = 0.5mm, draw = black, draw opacity = 1, thin, text opacity = 1, fill = black, fill opacity = 1},
  annotation/.style = {draw = black, draw opacity = 1, thin},
  ticks/.style = {draw = black, draw opacity = 1, thin},
  absorber/.style = {draw = black, draw opacity = 1, thin, pattern = north west lines},
  grid_line/.style = { very thin, dash pattern = on 1pt off 1pt, black },
  point_scatterer/.style = { fill = yellow, circle, inner sep = 1 pt },
  vector_lattice/.style = { ->, >={Latex[length=4pt,width=4pt]}, very thick, red },
  wavefront_plane/.style = { line width = 5 pt, RUBBLUE_RGB, opacity = 0.2 },
  wavefront_scattered/.style = { line width = 2 pt, RUBGREEN_RGB, opacity = 0.3 },
  vector_steering/.style = { ->, >={Latex[length=4pt,width=4pt]}, RUBBLUE_RGB },
  label_base/.style = { inner sep = 0, outer sep = 0, align = left },
  label_wavefront_incident/.style = { label_base, anchor = west }, 
  label_wavefronts_scattered/.style = { label_base, anchor = west }, 
  label_FOV/.style = { label_base, anchor = west }, 
  label_steering/.style = { label_base, anchor = west }, 
  pin_base/.style = { thin },
  pin_wavefront_incident/.style = { pin_base },
  pin_wavefronts_scattered/.style = { pin_base },
  pin_FOV/.style = { pin_base },
  pin_steering/.style = { pin_base },
]

\pgfmathsetmacro{\MinLengthX}{2.2}
\pgfmathsetmacro{\MaxLengthX}{2.2}

\pgfmathsetmacro{\AxisYStart}{0.2}
\pgfmathsetmacro{\AxisYStop}{-4.2}

\pgfmathsetmacro{\ElementTickHeight}{0.1}

\pgfmathsetmacro{\NElements}{16}
\pgfmathsetmacro{\ElementPitch}{0.25}
\pgfmathsetmacro{\ElementHeight}{0.4}
\pgfmathsetmacro{\ElementWidth}{0.2}

\pgfmathsetmacro{\PWPosZ}{-2}

\pgfmathsetmacro{\UVectThetaLength}{0.5}

%
\pgfmathsetmacro{\WidthElementIndex}{1}
\pgfmathsetmacro{\PitchElementIndex}{3}

\pgfmathsetmacro{\TickLength}{0.1}

\pgfmathsetmacro{\diameter}{0}

\pgfmathsetmacro{\MElements}{ ( \NElements - 1 ) / 2 }

\tikzset{
  pics/element/.style = {%
    code = {%
      \draw [ element, #1 ] (-\ElementWidth / 2, 0) rectangle (\ElementWidth / 2, \ElementHeight);
    }
  }
}

\fill [ black!10 ] (-\MinLengthX, 0) rectangle (\MaxLengthX, \AxisYStop);

\coordinate (object_function_pin) at ( -\MinLengthX + 0.25, \AxisYStop + 0.25 );
\node [ label_base, anchor = east ] (object_function_label) at ($(object_function_pin) - ( 0.5, 0 )$) { Object\\function };
\draw (object_function_label.east) -- (object_function_pin);

\coordinate (r1AxisMin) at (-\MinLengthX, 0);
\coordinate (r1AxisMax) at (\MaxLengthX, 0);
\coordinate (r2AxisMin) at (0, \AxisYStart);
\coordinate (r2AxisMax) at (0, \AxisYStop);

\draw [ axes_r ] (r1AxisMin) -- (r1AxisMax);	
\draw [ axes_r ] (r2AxisMin) -- (r2AxisMax);	

\foreach \x in {1,2,...,\NElements}{%

  \pgfmathsetmacro{\PosXCenter}{(\x - 1 - \MElements) * \ElementPitch}

  \pic at (\PosXCenter, 0) { element };

  \draw [thin] ( \PosXCenter, - \ElementTickHeight / 2 ) -- ++( 0, \ElementTickHeight );

} 

\node [ anchor = south west ] at ( { ( \NElements - 1 - \MElements ) * \ElementPitch + \ElementWidth / 2 }, \ElementHeight ) { Linear array };


\draw [ |-| ] ( { ( \PitchElementIndex - 1 - \MElements ) * \ElementPitch }, \ElementHeight + 0.2 )
  -- ( { ( \PitchElementIndex - \MElements ) * \ElementPitch }, \ElementHeight + 0.2 )
  node [ pos = 0.5, anchor = south ] { $p$ };

\begin{scope}

  \draw [ FOV ] ( -1.75, { - 0.5 * sqrt( 3 ) / 2 + 0.1 } ) rectangle ( 1.75, -4 );

  \coordinate (FOV_pin) at ( 1.75, -3.8 );
  \node [ label_FOV ] (label_FOV) at ($(FOV_pin) + ( 0.75, 0 )$) { Field of\\view };
  \draw [ pin_FOV ] (label_FOV) -- (FOV_pin);

  \clip ( -1.75, { - 0.5 * sqrt( 3 ) / 2 + 0.1 } ) rectangle ( 1.75, -4 );

  \begin{scope}[%
    cm={cos(0),sin(0),cos(-120),sin(-120),(0,0)},
  ]

    \draw [ grid_line ] ( -1.5, 0 ) grid[ step = 0.5 ] ( 4, 5 );

    \foreach \x in {-3,-2,...,8}{%
      \foreach \y in {1,2,...,9}{%
        \node[ point_scatterer ] (point-\x-\y) at ( 0.5 * \x, 0.5 * \y ) {};
      }
    }

    \coordinate (grid_pin) at ( 1, 3.5 );
    \draw [ vector_lattice ] (grid_pin) -- ++( 0.5, 0 )
      node [ pos = 0.6, anchor = south ] { $\vect{r}_{1}$ };
    \draw [ vector_lattice ] (grid_pin) -- ++( 0, 0.5 )
      node [ pos = 0.6, anchor = east ] { $\vect{r}_{2}$ };

  \end{scope}
  
\end{scope}

\coordinate (point_ref) at (point--1-3);
\node [ anchor = east, inner sep = 1pt, outer sep = 0, align = left ]  (label_ps) at ($(point_ref) + ( -1, 0.5 )$) { Point\\scatterers };
\draw (point_ref) -- (label_ps.east);

\node [ align = left, anchor = east ] (grid_label) at ($(grid_pin) - ( 1.5, - 0.25 )$) { Grid\\vectors };
\draw (grid_label.east) -- (grid_pin);

\draw [ wavefront_plane ] ( -2, \PWPosZ ) -- ( 2, \PWPosZ );
\draw [ wavefront_plane ] ( -2, \PWPosZ ) arc(-90:-135:0.25);
\draw [ wavefront_plane ] ( 2, \PWPosZ ) arc(-90:-45:0.25);

\draw [ vector_steering ] ( 1.9, \PWPosZ ) -- ++( 0, -\UVectThetaLength )
  coordinate [ pos = 0.5 ] (steering_pin)
  node [ pos = 1, anchor = north west, inner sep = 0, outer sep = 0 ] { $\uvect{\vartheta}$ };
\node [ label_steering ] (steering_label) at ($(steering_pin) + ( 0.6, -0.2 )$) { Direction };
\draw [ pin_steering ] (steering_label.west) -- (steering_pin);

\coordinate (wavefront_pin) at ( 2, \PWPosZ );
\node [ label_wavefront_incident ] (wavefront_label) at ($(wavefront_pin) + ( 0.5, 0.2 )$) { Steered\\plane wave };
\draw [ pin_wavefront_incident ] (wavefront_label.west) -- (wavefront_pin);

\begin{scope}
 
  \clip ( -\MinLengthX, 0 ) rectangle ( \MaxLengthX, \AxisYStop );

  \draw [ wavefront_scattered ] let
    \p1 = (point-2-2),
    \p2 = (point-4-3),
    \p3 = (point-4-4),
    \p4 = ( 0, \PWPosZ ),
    \n1 = { \y1 - \y4 },
    \n2 = { \y2 - \y4 },
    \n3 = { \y3 - \y4 } in
    (\p1) circle[ radius = \n1 ] coordinate (wavefront_scattered_pin_1) at ($(\p1) + (25:\n1)$)
    (\p2) circle[ radius = \n2 ] coordinate (wavefront_scattered_pin_2) at ($(\p2) + (25:\n2)$)
    (\p3) circle[ radius = \n3 ] coordinate (wavefront_scattered_pin_3) at ($(\p3) + (25:\n3)$);

\end{scope}

\node [ label_wavefronts_scattered ] (wavefronts_scattered_label) at ( 2.5, \PWPosZ + 1.0 ) { Scattered\\waves };
\draw [ pin_wavefronts_scattered ] (wavefronts_scattered_label.west) -- (wavefront_scattered_pin_1);
\draw [ pin_wavefronts_scattered ] (wavefronts_scattered_label.west) -- (wavefront_scattered_pin_2);
\draw [ pin_wavefronts_scattered ] (wavefronts_scattered_label.west) -- (wavefront_scattered_pin_3);

\node [ right ] at (r1AxisMax) { $x$ };
\node [ below ] at (r2AxisMax) { $z$ };

\end{tikzpicture}

%% file: theory/theory_spectral_properties.tex
%
\begin{figure}[t!]
 \centering%
  \input{theory/figures/latex/theory_passbands_pw.tex}
 \caption{}
 \label{fig:V}
\end{figure}
C
{
 The \ac{DAS} algorithm bandpass filters
 the object function.
 The algorithm, fixing both
 the wavenumber
 $k > 0$ and
 the steering angle
 $\vartheta \in ( -\pi / 2; \pi / 2 )$, recovers
 the object spectrum on
 \subref{fig:theory_passbands_pw_mono}
 the circular arc with
 the center
 $k \uvect{\vartheta}$ and
 the radius
 $k$.
 The angular aperture, which derives from
 the $F$-number, limits
 the receive angle
 $\varphi$.
 A sequence of
 $N = 3$ plane-wave measurements, varying
 the wavenumber
 $k$ within
 the interval
 $k \in [ k_{\text{lb}}; k_{\text{ub}} ]$, where
 $k_{\text{lb}} > 0$ and
 $k_{\text{ub}} > k_{\text{lb}}$ denote
 the lower and
 upper endpoints, respectively, recovers
 the object spectrum in
 \subref{fig:theory_passbands_pw_broadband_sequence}
 the union
 \eqref{eqn:theory_spectral_properties_support_pw_sequence} with
 the extremal angular spatial frequencies
 \eqref{eqn:theory_spectral_properties_support_pw_sequence_bounds}.
}%
{theory_passbands_pw}

Ultrasound images are
bandpass-filtered versions of
the acoustic object function.
This function describes
the spatial fluctuations in
the acoustic properties of
the object to be
imaged.
The bounded passbands, as will be shown in
\cref{subsec:theory_sampling}, enable
grids of
point scatterers to emulate
the object function in
the image formation.
The optimal grid parameters, however, depend on
the exact shapes of
the passbands.
This section, hence, will review
the passbands of
the images formed by
the \ac{DAS} algorithm in
\ac{CPWC}.
The maximum passbands, according to
the \acl{FDT}
\cite[Theorem 8.4]{book:Devaney2012},
\cite{proc:SchiffnerAI2012}, exclusively depend on
the steering angles and
the bandwidth of
the acquired \ac{RF} signals.
The array geometry,
the measurement noise, and
additional signal processing methods, such as
grating and
side lobe suppression, however, can reduce
the recoverable passbands. 

\subsubsection{Single Plane-Wave Images}
\input{theory/theory_spectral_properties_single.tex}

\subsubsection{Plane-Wave Compound Images}
\input{theory/theory_spectral_properties_sequence.tex}

%% file: theory/figures/latex/theory_passbands_pw.tex
%
\begin{tikzpicture}
[%
  font = \footnotesize,
  axes_k/.style = {->, thick},
  tick/.style = { thick, font = \scriptsize },
  union/.style = { black },
  spectra/.style = { inner color = white, outer color = black!80 },
  tx/.style = { orange },
  rx/.style = { cyan },
  sum/.style = { violet },
  ewald/.style = { draw opacity = 1, dash pattern = on 1pt off 1pt },
  ewald_tx/.style = { ewald, tx, very thick },
  ewald_rx/.style = { ewald, rx, very thick },
  vector/.style = { arrows = { -latex }, very thick },
  vector_tx/.style = { vector, tx },
  vector_rx/.style = { vector, rx },
  vector_sum/.style = { vector, sum },
  angle/.style = { ->, >=Stealth },
  angle_tx/.style = { angle, tx },
  angle_rx/.style = { angle, rx },
  vector_auxiliary/.style = { thin, dashed },
  annotation/.style = { thin, dashed, gray70 },
  title/.style = { inner sep = 0, outer sep = 0, above = 2ex, anchor = base west, font = \bfseries },
]

\pgfmathsetmacro{\lengthSemiaxis}{4}

\pgfmathsetmacro{\tickLength}{0.2}

\pgfmathsetmacro{\fnumber}{1}

\pgfmathsetmacro{\thetaRef}{ 50 }
\pgfmathsetmacro{\phiRef}{ atan( 1 / ( 2 * \fnumber ) ) }

\pgfmathsetmacro{\koRel}{0.4}

\pgfmathsetmacro{\thetaRadiusRel}{0.7}
\pgfmathsetmacro{\phiRadiusRel}{0.7}

\pgfmathsetmacro{\thetao}{-20}
\pgfmathsetmacro{\thetai}{0}
\pgfmathsetmacro{\thetaii}{10}

\pgfmathsetmacro{\klboRel}{0.1}
\pgfmathsetmacro{\klbiRel}{0.1}
\pgfmathsetmacro{\klbiiRel}{0.1}

\pgfmathsetmacro{\kuboRel}{0.35}
\pgfmathsetmacro{\kubiRel}{0.35}
\pgfmathsetmacro{\kubiiRel}{0.35}

\pgfmathsetmacro{\phiLB}{ - atan( 1 / ( 2 * \fnumber ) ) }
\pgfmathsetmacro{\phiUB}{ - \phiLB }

\def\configListPWs{-20/\klboRel/\kuboRel/RUBGRAY_RGB,0/\klbiRel/\kubiRel/RUBBLUE_RGB,10/\klbiiRel/\kubiiRel/yellow}

\pgfmathsetmacro{\kxAxisStart}{-\lengthSemiaxis}
\pgfmathsetmacro{\kxAxisStop}{\lengthSemiaxis}
\pgfmathsetmacro{\kyAxisStart}{ \lengthSemiaxis / 10 }
\pgfmathsetmacro{\kyAxisStop}{ - \lengthSemiaxis }

\pgfmathsetmacro{\ko}{ \koRel * \lengthSemiaxis }

\pgfmathsetmacro{\thetaRadius}{ \thetaRadiusRel * \ko }

\pgfmathsetmacro{\phiRadius}{ \phiRadiusRel * \ko }

\pgfmathsetmacro{\klbo}{ \klboRel * \lengthSemiaxis }
\pgfmathsetmacro{\klbi}{ \klbiRel * \lengthSemiaxis }
\pgfmathsetmacro{\klbii}{ \klbiiRel * \lengthSemiaxis }

\pgfmathsetmacro{\kubo}{ \kuboRel * \lengthSemiaxis }
\pgfmathsetmacro{\kubi}{ \kubiRel * \lengthSemiaxis }
\pgfmathsetmacro{\kubii}{ \kubiiRel * \lengthSemiaxis }

\pgfmathsetmacro{\SpectrumRadius}{ sqrt( pow( \kxAxisStop, 2 ) + pow( \kyAxisStop, 2 ) }

\pgfmathsetmacro{\PWklatlb}{ \kubii * ( sin( \thetao ) - sin( \phiUB ) ) }
\pgfmathsetmacro{\PWklatub}{ \kubii * ( sin( \thetaii ) + sin( \phiUB ) ) }
\pgfmathsetmacro{\PWkaxlb}{ \klbo * ( cos( \thetao ) + cos( \phiUB ) }
\pgfmathsetmacro{\PWkaxub}{ \kubii * ( cos( \thetai ) + 1 ) }

\newcommand{\drawaxes}{%

  \draw [axes_k] (\kxAxisStart, 0) -- (\kxAxisStop, 0);	
  \draw [axes_k] (0, \kyAxisStart) -- (0, \kyAxisStop);	

  \node [above, anchor = south] at (\kxAxisStop, 0) { $\hat{k}_{x}$ };
  \node [right] at (0, \kyAxisStop) { $\hat{k}_{z}$ };

}

\newcommand{\arcThroughThreePoints}[4]{
\draw[ #1 ]
  let \p0 = ($(#3) - (#2)$),
      \p1 = ($(#4) - (#2)$),
      \n0 = { veclen( \p0 ) },	 
      \n1 = { -acos( \x0 / \n0 ) }, 
      \n2 = { -acos( \x1 / \n0 ) }  
  in (#3) arc(\n1:\n2:\n0);
}

\newcommand{\drawEnvelopeUnion}[3]{%

  \path [red, opacity=0.5, name intersections={of = #1 and #2, name = i01, total = \t}];
  \path [red, opacity=0.5, name intersections={of = #2 and #3, name = i12, total = \t}];
  \path [red, opacity=0.5, name intersections={of = #1 and #3, name = i02, total = \t}];

  \arcThroughThreePoints{union}{-110:\kubo}{$(-110:\kubo) + (-90 + \phiLB:\kubo)$}{i01-4}
  \draw [ union ] ($(-110:\kubo) + (-90 + \phiLB:\kubo)$) -- ($(-110:\klbo) + (-90 + \phiLB:\klbo)$);
  \arcThroughThreePoints{union}{-110:\klbo}{$(-110:\klbo) + (-90 + \phiLB:\klbo)$}{i01-2}
  \draw [ union ] (i01-2) -- ($(-90:\klbi) + (-90 + \phiLB:\klbi)$);

  \arcThroughThreePoints{union}{-80:\kubii}{i12-4}{$(-80:\kubii) + (-90 + \phiUB:\kubii)$}
  \draw [ union ] ($(-80:\kubii) + (-90 + \phiUB:\kubii)$) -- ($(-80:\klbii) + (-90 + \phiUB:\klbii)$);
  \arcThroughThreePoints{union}{-90:\klbi}{$(-90:\klbi) + (-90 + \phiLB:\klbi)$}{i12-2}
  \draw [ union ] (i12-2) -- ($(-80:\klbii) + (-90 + \phiLB:\klbii)$);
  \arcThroughThreePoints{union}{-80:\klbii}{$(-80:\klbii) + (-90 + \phiUB:\klbii)$}{i12-3}
  \draw [ union ] (i12-3) -- ($(-90:\klbi) + (-90 + \phiUB:\klbi)$);
  \arcThroughThreePoints{union}{-90:\klbi}{$(-90:\klbi) + (-90 + \phiUB:\klbi)$}{i12-1}
  \arcThroughThreePoints{union}{-80:\klbii}{i12-1}{i02-3}

  \arcThroughThreePoints{union}{-90:\kubi}{i01-4}{i12-4}
  \arcThroughThreePoints{union}{-80:\klbii}{$(-80:\klbii) + (-90 + \phiLB:\klbii)$}{i02-1}
  \arcThroughThreePoints{union}{-110:\klbo}{i02-1}{$(-110:\klbo) + (-90 + \phiUB:\klbo)$}
  \draw [ union ] ($(-110:\klbo) + (-90 + \phiUB:\klbo)$) -- (i01-3);

}

\begin{scope}

  \begin{scope}
    \clip (\kxAxisStart, 0) rectangle (\kxAxisStop, \kyAxisStop);
    \shade [ spectra ] (0,0) circle [ radius = \SpectrumRadius ];
  \end{scope}

  \drawaxes

  \draw [ tick, tx ] (-\ko, -\tickLength / 2) -- (-\ko, \tickLength / 2)
    node [ above = 1ex, anchor = base ] { $-k$ };
  \draw [ tick, tx ] (\ko, -\tickLength / 2) -- (\ko, \tickLength / 2)
    node [ above = 1ex, anchor = base ] { $k$ };

  \draw [ tick, tx ] ( -\tickLength / 2, -\ko ) -- (\tickLength / 2, -\ko)
    node [ anchor = north east ] { $k$ };

  \node [ inner sep = 0, outer sep = 0, anchor = north east, align = left ] (labelFTs) at ($(-170:0.2*\SpectrumRadius) + (-0.5, -0.3)$)
    { Object\\spectrum };
  \draw (-170:0.2*\SpectrumRadius) -- (labelFTs.north east);

  \draw [ ewald_tx ] ( -\ko, 0 ) arc(-180:0:\ko);

  \draw [ ewald_rx ] ($(-90 + \thetaRef:\ko) + (-90 + \phiLB:\ko)$) arc(-90 + \phiLB:-90 + \phiUB:\ko)
    node [ pos = 0.5, anchor = north east, font = \scriptsize, align = left, inner sep = 1 pt, outer sep = 0 ] { Angular\\aperture };

  \coordinate (vector_tx_end) at (-90 + \thetaRef:\ko);
  \draw [ vector_tx ] ( 0, 0 ) -- (vector_tx_end)
    node [ pos = 0.5, anchor = south west, font = \scriptsize, inner sep = 1pt, outer sep = 0 ] { $k \uvect{\vartheta}$ };

  \draw [ angle_tx ] (-90:\thetaRadius) arc(-90:-90 + \thetaRef:\thetaRadius)
    node [ pos = 0.4, anchor = north, font = \scriptsize ] { $\vartheta$ };

  \draw [ vector_rx ] (vector_tx_end) -- ++(-90 + \phiRef:\ko)
    node [ pos = 0.5, anchor = south west, font = \scriptsize, inner sep = 1pt, outer sep = 0 ] { $k \uvect{\varphi}$ };
  \draw [ rx ] (vector_tx_end) -- ++(-90 + \phiLB:\ko);

  \draw [ rx, dash pattern = on 1pt off 1pt ] (vector_tx_end) -- ($(vector_tx_end) + (-90:\ko)$);
  \draw [ angle_rx ] ($(vector_tx_end) + (-90:\phiRadius)$) arc(-90:-90 + \phiRef:\phiRadius)
    node [ pos = 0.4, anchor = north, font = \scriptsize ] { $\varphi$ };

  \draw [ vector_sum ] ( 0, 0 ) -- ($(vector_tx_end)+(-90 + \phiRef:\ko)$)
    node [ anchor = north west, inner sep = 1pt, outer sep = 0 ] { $\vect{T}( k, \vartheta, \varphi )$ };

  \node [ title ] at ( \kxAxisStart, \kyAxisStart )
    { \phantomsubcaption\label{fig:theory_passbands_pw_mono}(a) Monofrequent steered plane wave };

\end{scope}

\begin{scope}[yshift = -5.5cm]

  \begin{scope}
    \clip (\kxAxisStart, 0) rectangle (\kxAxisStop, \kyAxisStop);
    \shade [ spectra ] (0,0) circle [ radius = \SpectrumRadius ];
  \end{scope}

  \drawaxes

  \draw [ tick ] (\PWklatlb, -\tickLength / 2) -- (\PWklatlb, \tickLength / 2) node [ above = 1ex, anchor = base ] { $\hat{k}_{x, \text{lb}}$ };
  \draw [ tick ] (\PWklatub, -\tickLength / 2) -- (\PWklatub, \tickLength / 2) node [ above = 1ex, anchor = base ] { $\hat{k}_{x, \text{ub}}$ };

  \draw [ tick ] (-\tickLength / 2, -\PWkaxlb) -- (\tickLength / 2, -\PWkaxlb) node [ anchor = base west ] { $\hat{k}_{z, \text{lb}}$ };
  \draw [ tick ] (-\tickLength / 2, -\PWkaxub) -- (\tickLength / 2, -\PWkaxub) node [ anchor = north west ] { $\hat{k}_{z, \text{ub}}$ };

  \node [ inner sep = 0, outer sep = 0, anchor = north east, align = left ] (labelFTs) at ($(-170:0.2*\SpectrumRadius) + (-0.5, -0.3)$)
    { Object\\spectrum };
  \draw (-170:0.2*\SpectrumRadius) -- (labelFTs.north east);

  \foreach \theta / \klbRel / \kubRel / \col [ count = \indexN from 0 ] in \configListPWs
  {
    \pgfmathsetmacro{\klbact}{ \klbRel * \lengthSemiaxis }
    \pgfmathsetmacro{\kubact}{ \kubRel * \lengthSemiaxis }

    \ifthenelse{\theta < 0}{%

      \filldraw[ name path global=setK-\indexN, color = \col, dash pattern = on 1pt off 1pt, draw opacity = 1, fill opacity = 0.15 ]
           ($(-90 + \theta:\klbact) + (-90 + \phiLB:\klbact)$) arc(-90 + \phiLB:-90 + \phiUB:\klbact)
        -- ($(-90 + \theta:\kubact) + (-90 + \phiUB:\kubact)$) arc(-90 + \phiUB:-90 + \phiLB:\kubact)
           coordinate [ pos = 0.5 ] (pos-\indexN)
        -- cycle coordinate [ pos = 0.5 ] (label-\indexN);

      \node [ inner sep = 0, outer sep = 0, anchor = east, align = left, \col, font = \tiny ] (label-\indexN-node) at ($(label-\indexN) - (0.3, 0.3)$)
        { Passband $\hat{K}( \SI{\theta}{\degree} )$ };
      \draw [ \col ] (label-\indexN) -- (label-\indexN-node.east);

    }{%

      \filldraw[ name path global=setK-\indexN, color = \col, dash pattern = on 1pt off 1pt, draw opacity = 1, fill opacity = 0.15 ]
           ($(-90 + \theta:\klbact) + (-90 + \phiLB:\klbact)$) arc(-90 + \phiLB:-90 + \phiUB:\klbact)
        -- ($(-90 + \theta:\kubact) + (-90 + \phiUB:\kubact)$)
           coordinate [ pos = 0.5 ] (label-\indexN) arc(-90 + \phiUB:-90 + \phiLB:\kubact)
        -- cycle;

      \node [ inner sep = 0, outer sep = 0, anchor = west, align = left, \col, font = \tiny ] (label-\indexN-node) at ($(label-\indexN) + ( 0.3, { ( \indexN - 2 ) * 0.3 } )$)
        { Passband $\hat{K}( \SI{\theta}{\degree} )$ };
      \draw [ \col ] (label-\indexN) -- (label-\indexN-node.west);

    }
  }

  \drawEnvelopeUnion{setK-0}{setK-1}{setK-2}

  \node [ inner sep = 0, outer sep = 0, anchor = north east, align = left, font = \scriptsize ] (labelUnion) at ($(pos-0) - (0.5, 0.5)$) { Total\\passband };
  \draw (pos-0) -- (labelUnion.north east);

  \node [ title ] at ( \kxAxisStart, \kyAxisStart )
    { \phantomsubcaption\label{fig:theory_passbands_pw_broadband_sequence}(b) Sequence of three steered plane waves };

\end{scope}

\end{tikzpicture}

%% file: theory/theory_spectral_properties_single.tex
A single plane-wave measurement enables
the formation of
a low-quality image.
The recoverable passband
$\hat{K}( \vartheta )$ for
a \ac{PW} with
the steering angle
$\vartheta \in ( -\pi / 2; \pi / 2 )$ equals
\begin{subequations}
\label{eqn:theory_spectral_properties_support_pw_single}
\begin{equation}
\begin{split}
  \hat{K}( \vartheta )
  =
  \Bigl\{
  & \hat{\vect{k}} \in \R^{2}:
    \hat{\vect{k}} = \vect{T}( k, \vartheta, \varphi ),\\
  & ( k, \varphi ) \in [ k_{\text{lb}}; k_{\text{ub}} ] \times [ \varphi_{\text{lb}}; \varphi_{\text{ub}} ]
  \Bigr\},
\end{split}
\end{equation}
where
$\vect{T}$ is
the coordinate transform
\begin{equation}
  \vect{T}( k, \vartheta, \varphi )
  =
  k
  \left(
    \uvect{\vartheta}
  + \uvect{\varphi}
  \right)
\end{equation}
\end{subequations}
with
the wavenumber
$k > 0$,
the receive angle
$\varphi \in ( -\pi / 2; \pi / 2 )$, and
the unit vectors
$\uvect{\vartheta} = \trans{ ( \sin( \vartheta ), \cos( \vartheta ) ) }$ and
$\uvect{\varphi} = \trans{ ( \sin( \varphi ), \cos( \varphi ) ) }$.
This transform, for
a given steering angle
$\vartheta$, maps
the wavenumber
$k$ and
the receive angle
$\varphi$ to
the angular spatial frequencies
$\hat{\vect{k}} = \trans{ ( \hat{k}_{x}, \hat{k}_{z} ) }$.
These frequencies, fixing
the wavenumber
$k$, form
a circular arc with
the center
$k \uvect{\vartheta}$ and
the radius
$k$, as shown in
\cref{fig:theory_passbands_pw_mono}.
The angular aperture of
this arc derives from
the $F$-number
$F > 0$ in
the receive focusing
\cite{proc:SchiffnerIUS2021} and imposes
the lower and
upper bounds
$\varphi_{\text{lb}} = - \varphi_{\text{ub}}$ and
$\varphi_{\text{ub}} = \arctan(  1 / ( 2 F ) )$,
respectively, on
the receive angle
$\varphi$.
The wavenumber
$k$ has
the lower and
upper bounds
$k_{\text{lb}} > 0$ and
$k_{\text{ub}} > k_{\text{lb}}$,
respectively, because of
the limited bandwidth of
the acquired \ac{RF} signals.

%% file: theory/theory_spectral_properties_sequence.tex
The superposition of
the low-quality images obtained from
a sequence of
$N \in \N$ plane-wave measurements improves
the image quality at
the expense of
the frame rate.
The recoverable passbands
\eqref{eqn:theory_spectral_properties_support_pw_single} for
the steering angles
$\vartheta_{0} < \vartheta_{1} < \dotsb < \vartheta_{N - 1}$, as shown in
\cref{fig:theory_passbands_pw_broadband_sequence}, merge into
the total recoverable passband
\begin{equation}
  \hat{K}
  =
  \bigcup_{ n = 0 }^{ N - 1 }
    \hat{K}( \vartheta_{n} ).
 \label{eqn:theory_spectral_properties_support_pw_sequence}
\end{equation}
This enlarged passband explains
the higher image quality and defines
the bounds on
the angular spatial frequencies
\begin{subequations}
\label{eqn:theory_spectral_properties_support_pw_sequence_bounds}
\begin{align}
  \hat{k}_{x, \text{lb}}
  &=
  k_{\text{ub}}
  \bigl[
    \sin( \vartheta_{0} )
    -
    \sin( \varphi_{\text{ub}} )
  \bigr],\\
  \hat{k}_{x, \text{ub}}
  &=
  k_{\text{ub}}
  \bigl[
    \sin( \vartheta_{N - 1} )
    +
    \sin( \varphi_{\text{ub}} )
  \bigr],\\
  \hat{k}_{z, \text{lb}}
  &=
  k_{\text{lb}}
  \bigl[
    \min\bigl\{ \cos( \vartheta_{0} ), \cos( \vartheta_{N - 1} ) \bigr\}
    +
    \cos( \varphi_{\text{ub}} )
  \bigr],\\
  \hat{k}_{z, \text{ub}}
  &=
  k_{\text{ub}}
  \bigl[
    \underset{n}{\max}\bigl\{ \cos( \vartheta_{n} ) \bigr\}
    +
    1
  \bigr],
\end{align}
\end{subequations}
which will now be used in
the optimization of
the grid parameters.

%% file: theory/theory_sampling.tex
%
\begin{figure}[t!]
 \centering%
  \input{theory/figures/latex/theory_sampling_pw.tex}
 \caption{}
 \label{fig:V}
\end{figure}
C
{
 Effect of
 the discretization of
 the bandpass-filtered object function.
 The discretization superimposes
 copies of
 the object spectrum in
 the spatial Fourier domain.
 These copies are limited to
 the total recoverable passband
 \eqref{eqn:theory_spectral_properties_support_pw_sequence} and arise on
 the regular grid
 \eqref{eqn:theory_sampling_grid_spectrum}.
 The bandpass-filtered object function and, thus,
 the ultrasound image can be recovered from
 this superposition if
 the grid vectors
 $\vect{u}_{1}$ and
 $\vect{u}_{2}$ prevent
 the copies from
 overlapping.
 Given
 suitable spectral grid vectors
 $\vect{u}_{1}$ and
 $\vect{u}_{2}$,
 the spatial grid vectors
 $\vect{r}_{1}$ and
 $\vect{r}_{2}$ equal
 \eqref{eqn:theory_sampling_components}.
}%
{theory_sampling_pw}

The image, as explained in
\cref{subsec:theory_spectral_properties}, at best equals
the bandpass-filtered object function.
The spectrum of
this function equals
the spectrum of
the object function inside
the total recoverable passband
\eqref{eqn:theory_spectral_properties_support_pw_sequence} but
zero elsewhere.
Such a function, according to
the bivariate sampling theorem
\cite{article:PetersenInfCtrl1962}, can be recovered from
its samples on
the regular grid
\begin{equation}
  \setsymbol{G}
  =
  \left\{
    \vect{r} \in \R^{2}:
    \vect{r} = l_{1} \vect{r}_{1} + l_{2} \vect{r}_{2},
    l_{1}, l_{2} \in \Z
  \right\},
 \label{eqn:theory_sampling_grid_space}
\end{equation}
where
$\Z$ is
the set of
all integers and
$\vect{r}_{1}$ and
$\vect{r}_{2}$ are
linearly independent vectors that will now be determined. 

The sampling superimposes
copies of
the bandpass-filtered object spectrum in
the spatial Fourier domain, as shown in
\cref{fig:theory_sampling_pw}.
These copies arise at
all points of
the regular grid 
\begin{equation}
  \setsymbol{H}
  =
  \left\{
    \vect{u} \in \R^{2}:
    \vect{u} = m_{1} \vect{u}_{1} + m_{2} \vect{u}_{2},
    m_{1}, m_{2} \in \Z
  \right\},
 \label{eqn:theory_sampling_grid_spectrum}
\end{equation}
where
the dot products of
the vectors
$\vect{u}_{1}$ and
$\vect{u}_{2}$ with
the vectors
$\vect{r}_{1}$ and
$\vect{r}_{2}$ satisfy
$\inprod{ \vect{u}_{l} }{ \vect{r}_{m} } = 2 \pi \kroneckerdelta{l}{m}$ with
the Kronecker delta
$\kroneckerdelta{l}{m}$ for
all $l, m \in \{ 1, 2 \}$
\cite{article:PetersenInfCtrl1962}.
The bandpass-filtered object function and, thus,
the ultrasound image can be recovered from
the superposition if
the vectors
$\vect{u}_{1}$ and
$\vect{u}_{2}$ prevent
the copies from overlapping.
%
The vectors
$\vect{r}_{1}$ and
$\vect{r}_{2}$, given
suitable vectors
$\vect{u}_{1} = \trans{ ( u_{1,x}, u_{1,z} ) }$ and
$\vect{u}_{2} = \trans{ ( u_{2,x}, u_{2,z} ) }$, then equal
\begin{align}
  \vect{r}_{1}
  &=
  \frac{ 2 \pi }{ D }
  \begin{pmatrix}
    u_{2,z}\\
  - u_{2,x}
  \end{pmatrix}
  & & \text{and} &
  \vect{r}_{2}
  &=
  \frac{ 2 \pi }{ D }
  \begin{pmatrix}
  - u_{1,z}\\
    u_{1,x}
  \end{pmatrix}
 \label{eqn:theory_sampling_components}
\end{align}
with
$D = u_{1,x} u_{2,z} - u_{1,z} u_{2,x}$.
Two spatial sampling grids
\eqref{eqn:theory_sampling_grid_space} that enable
the exact recovery of
the ultrasound image will now be derived.

\subsubsection{Optimal Orthogonal Grid}
\label{subsubsec:theory_sampling_orthogonal}
\input{theory/theory_sampling_orthogonal.tex}

\subsubsection{Proposed Rhombic Grid}
\input{theory/theory_sampling_rhombic.tex}

%% file: theory/figures/latex/theory_sampling_pw.tex
%
\begin{tikzpicture}
[%
  font = \footnotesize,
  axes_k/.style = {->, thick},
  tick/.style = { thick, font = \scriptsize },
  union/.style = { black },
  spectra/.style = { inner color = white, outer color = black!80 },
  vector_lattice/.style = { ->, >=Latex, very thick, red },
  help_lines/.style = { dashed },
]

\pgfmathsetmacro{\lengthSemiaxis}{4}

\pgfmathsetmacro{\AxisLengthX}{3.0}   
\pgfmathsetmacro{\AxisLengthY}{3.0}   

\pgfmathsetmacro{\ColorbarLengthX}{0.2}                 
\pgfmathsetmacro{\ColorbarLengthY}{ \AxisLengthY }      
\pgfmathsetmacro{\ColorbarDistanceX}{0.2}               

\pgfmathsetmacro{\tickLength}{0.2}

\pgfmathsetmacro{\fnumber}{1}

\pgfmathsetmacro{\thetao}{-80}

\pgfmathsetmacro{\klboRel}{0.1}
\pgfmathsetmacro{\klbiRel}{0.1}
\pgfmathsetmacro{\klbiiRel}{0.1}

\pgfmathsetmacro{\kuboRel}{0.35}
\pgfmathsetmacro{\kubiRel}{0.35}
\pgfmathsetmacro{\kubiiRel}{0.35}

\pgfmathsetmacro{\vecUOneAngle}{-30}
\pgfmathsetmacro{\vecuTwoAngle}{-90}

\pgfmathsetmacro{\AxisLimitsXLBCIRS}{-19.25}  
\pgfmathsetmacro{\AxisLimitsXUBCIRS}{19.25}   
\pgfmathsetmacro{\AxisLimitsZLBCIRS}{5.4}     
\pgfmathsetmacro{\AxisLimitsZUBCIRS}{43.9}    

\pgfmathsetmacro{\ImageLimitsKXLBCIRS}{-0.5010} 
\pgfmathsetmacro{\ImageLimitsKXUBCIRS}{0.4990}  
\pgfmathsetmacro{\ImageLimitsKZLBCIRS}{-0.001}   
\pgfmathsetmacro{\ImageLimitsKZUBCIRS}{0.999}  

\pgfmathsetmacro{\AxisLimitsKXLBCIRS}{-0.5010}  
\pgfmathsetmacro{\AxisLimitsKXUBCIRS}{0.4990}   
\pgfmathsetmacro{\AxisLimitsKZLBCIRS}{-0.001}     
\pgfmathsetmacro{\AxisLimitsKZUBCIRS}{0.999}    

\pgfmathsetmacro{\AxisLabelXShift}{-0.15}  
\pgfmathsetmacro{\AxisLabelZShift}{-0.18}  

\pgfmathsetmacro{\phiLB}{ - atan( 1 / ( 2 * \fnumber ) ) }
\pgfmathsetmacro{\phiUB}{ - \phiLB }

\def\configListPWs{-20/\klboRel/\kuboRel,0/\klbiRel/\kubiRel,10/\klbiiRel/\kubiiRel}

\pgfmathsetmacro{\kxAxisStart}{ - \lengthSemiaxis }
\pgfmathsetmacro{\kxAxisStop}{ \lengthSemiaxis }
\pgfmathsetmacro{\kyAxisStart}{ 0.1 * \lengthSemiaxis }
\pgfmathsetmacro{\kyAxisStop}{ - \lengthSemiaxis }

\pgfmathsetmacro{\klbo}{ \klboRel * \lengthSemiaxis }
\pgfmathsetmacro{\klbi}{ \klbiRel * \lengthSemiaxis }
\pgfmathsetmacro{\klbii}{ \klbiiRel * \lengthSemiaxis }

\pgfmathsetmacro{\kubo}{ \kuboRel * \lengthSemiaxis }
\pgfmathsetmacro{\kubi}{ \kubiRel * \lengthSemiaxis }
\pgfmathsetmacro{\kubii}{ \kubiiRel * \lengthSemiaxis }

\pgfmathsetmacro{\SpectrumRadius}{ sqrt( \kxAxisStop * \kxAxisStop + \kyAxisStop * \kyAxisStop ) }

\pgfmathsetmacro{\PWkaxlb}{ \klbo * ( cos( -20 ) + cos( \phiUB ) ) }
\pgfmathsetmacro{\PWkaxub}{ \kubi * ( cos( 0 ) + 1 ) }

\pgfmathsetmacro{\vecUOneLat}{ cos( \vecUOneAngle ) }
\pgfmathsetmacro{\vecUOneAx}{ sin( \vecUOneAngle ) }

\pgfmathsetmacro{\vecUTwoLat}{ cos( \vecuTwoAngle ) }
\pgfmathsetmacro{\vecUTwoAx}{ sin( \vecuTwoAngle ) }

\pgfmathsetmacro{\vecUOneLength}{ \PWkaxub - \PWkaxlb }
\pgfmathsetmacro{\vecUTwoLength}{ \vecUOneLength }

\newcommand{\drawaxes}{%

  \draw [axes_k] (\kxAxisStart, 0) -- (\kxAxisStop, 0);	
  \draw [axes_k] (0, \kyAxisStart) -- (0, \kyAxisStop);	

  \node [ anchor = south ] at (\kxAxisStop, 0) { $\hat{k}_{x}$ };
  \node [ anchor = north west ] at (0, \kyAxisStop) { $\hat{k}_{z}$ };

}

\begin{scope}

  \begin{scope}
    \fill [ black ] (\kxAxisStart, 0) rectangle (\kxAxisStop, \kyAxisStop); 
    \clip (\kxAxisStart, 0) rectangle (\kxAxisStop, \kyAxisStop); 
    \foreach \shiftUOne in {-2,-1,0,1,2}
      \foreach \shiftUTwo in {-2,-1,0,1,2}
      {
        \coordinate (shiftact) at ( { \shiftUOne * \vecUOneLat * \vecUOneLength + \shiftUTwo * \vecUTwoLat * \vecUTwoLength }, { \shiftUOne * \vecUOneAx * \vecUOneLength + \shiftUTwo * \vecUTwoAx * \vecUTwoLength } );
        \foreach \theta / \klbRel / \kubRel [ count = \indexN from 0 ] in \configListPWs
        {
          \pgfmathsetmacro{\klbact}{ \klbRel * \lengthSemiaxis }
          \pgfmathsetmacro{\kubact}{ \kubRel * \lengthSemiaxis }

          \begin{scope}
            \clip
                 ($(-90 + \theta:\klbact) + (-90 + \phiLB:\klbact) + (shiftact)$) arc(-90 + \phiLB:-90 + \phiUB:\klbact)
              -- ($(-90 + \theta:\kubact) + (-90 + \phiUB:\kubact) + (shiftact)$) arc(-90 + \phiUB:-90 + \phiLB:\kubact)
              -- cycle;
            \shade [ spectra ] (shiftact) circle [ radius = \SpectrumRadius ];
          \end{scope}
        }
      }
    \foreach \shiftUTwo in {-2,-1,0,1,2}
    {
      \coordinate (left) at ( - 3 * \vecUOneLat * \vecUOneLength, { - 3 * \vecUOneAx * \vecUOneLength + \shiftUTwo * \vecUTwoAx * \vecUTwoLength } );
      \coordinate (right) at ( 3 * \vecUOneLat * \vecUOneLength, { + 3 * \vecUOneAx * \vecUOneLength + \shiftUTwo * \vecUTwoAx * \vecUTwoLength } );
      \draw [ help_lines ] (left) -- (right);
    }
    \foreach \shiftUOne in {-2,-1,0,1,2}
    {
      \coordinate (top) at ( \shiftUOne * \vecUOneLat * \vecUOneLength, { \shiftUOne * \vecUOneAx * \vecUOneLength + 3 * \vecUTwoAx * \vecUTwoLength } );
      \coordinate (bottom) at ( \shiftUOne * \vecUOneLat * \vecUOneLength, { \shiftUOne * \vecUOneAx * \vecUOneLength - 3 * \vecUTwoAx * \vecUTwoLength } );
      \draw [ help_lines ] (top) -- (bottom);
    }
  \end{scope}

  \drawaxes

  \draw [ vector_lattice ]
    ( - \vecUOneLat * \vecUOneLength + \vecUTwoLat * \vecUTwoLength, - \vecUOneAx * \vecUOneLength + \vecUTwoAx * \vecUTwoLength )
  -- ++ ( \vecUOneLat * \vecUOneLength, \vecUOneAx * \vecUOneLength )
    node [ pos = 0.6, anchor = south ] { $\vect{u}_{1}$ };
  \draw [ vector_lattice ]
    ( - \vecUOneLat * \vecUOneLength + \vecUTwoLat * \vecUTwoLength, - \vecUOneAx * \vecUOneLength + \vecUTwoAx * \vecUTwoLength )
  -- ++ ( \vecUTwoLat * \vecUTwoLength, \vecUTwoAx * \vecUTwoLength )
    node [ pos = 0.6, anchor = west ] { $\vect{u}_{2}$ };
  \node [ anchor = east, font = \scriptsize, red, align = left ] at ( - \vecUOneLat * \vecUOneLength + \vecUTwoLat * \vecUTwoLength, - \vecUOneAx * \vecUOneLength + \vecUTwoAx * \vecUTwoLength )
    { Grid\\vectors };

\end{scope}

\end{tikzpicture}

%% file: theory/theory_sampling_orthogonal.tex
The spectral grid vectors
$\vect{u}_{1}$ and
$\vect{u}_{2}$ prevent
the copies of
the bandpass-filtered object spectrum in
\cref{fig:theory_sampling_pw} from
overlapping if
$\vect{u}_{1} = ( \hat{k}_{x, \text{ub}} - \hat{k}_{x, \text{lb}} ) \uvect{x}$ and
$\vect{u}_{2} = ( \hat{k}_{z, \text{ub}} - \hat{k}_{z, \text{lb}} ) \uvect{z}$, where
$\hat{k}_{x, \text{lb}}$,
$\hat{k}_{x, \text{ub}}$,
$\hat{k}_{z, \text{lb}}$,
$\hat{k}_{z, \text{ub}}$ are
the bounds
\eqref{eqn:theory_spectral_properties_support_pw_sequence_bounds} and
$\uvect{x} = \trans{ ( 1, 0 ) }$ and
$\uvect{z} = \trans{ ( 0, 1 ) }$ are
orthonormal unit vectors.
This choice results in
an \emph{orthogonal} sampling grid
\eqref{eqn:theory_sampling_grid_space}, and
the spatial grid vectors
\eqref{eqn:theory_sampling_components} become
$\vect{r}_{1} = \Delta x \uvect{x}$ and
$\vect{r}_{2} = \Delta z \uvect{z}$ with
the optimal lengths
\begin{align}
  \Delta x
  &=
  \frac{ 2 \pi }{ \hat{k}_{x, \text{ub}} - \hat{k}_{x, \text{lb}} }
  & & \text{and} &
  \Delta z
  &=
  \frac{ 2 \pi }{ \hat{k}_{z, \text{ub}} - \hat{k}_{z, \text{lb}} }.
 \label{eqn:theory_sampling_orthogonal_spacing}
\end{align}

%% file: theory/theory_sampling_rhombic.tex
The spectral grid vectors
$\vect{u}_{1} = \Delta u \trans{ ( \sqrt{3} / 2, 1 / 2 ) }$ and
$\vect{u}_{2} = \Delta u \uvect{z}$ also prevent
the copies of
the bandpass-filtered object spectrum in
\cref{fig:theory_sampling_pw} from
overlapping if
$\Delta u = \hat{k}_{z, \text{ub}} - \hat{k}_{z, \text{lb}}$, where
$\hat{k}_{z, \text{lb}}$ and
$\hat{k}_{z, \text{ub}}$ are
the bounds
\eqref{eqn:theory_spectral_properties_support_pw_sequence_bounds} and
$\uvect{z} = \trans{ ( 0, 1 ) }$.
This choice results in
a \SI{120}{\degree} rhombic grid
\eqref{eqn:theory_sampling_grid_space}, which is shown in
\cref{fig:theory_grid}, and
the spatial grid vectors
\eqref{eqn:theory_sampling_components} become
\begin{align}
  \vect{r}_{1}
  &=
  \Delta r
  \uvect{x}
  & & \text{and} &
  \vect{r}_{2}
  &=
  \Delta r
  \begin{pmatrix}
  - 1 / 2\\
    \sqrt{3} / 2
  \end{pmatrix}
 \label{eqn:theory_sampling_rhombic_vectors}
\end{align}
with
the length
$\Delta r = 4 \pi / ( \sqrt{3} \Delta u )$.

%% file: methods/methods.tex
The advantages of
the proposed rhombic grid
\eqref{eqn:theory_sampling_rhombic_vectors} over
two orthogonal grids, which served as
benchmarks, were confirmed in
an experiment with
a commercial multi-tissue phantom%
\footnote{%
  Computerized Imaging Reference Systems (CIRS), Inc., Norfolk, VA, USA%
  \label{ftnote:manufacturer_cirs}
}
(%
  model: 040;
  average speed of sound: $c = \SI{1538.75}{\meter\per\second}$%
).
A SonixTouch Research system%
\footnote{%
  Analogic Corporation, Sonix Design Center, Richmond, BC, Canada
  \label{ftnote:manufacturer_analogic}
} with
a linear array
(%
  model: L14-5/38;
  number of elements: $N_{\text{el}} = 128$,
  pitch: $p = \SI{304.8}{\micro\meter}$%
) acquired and stored
the \ac{RF} signals induced by
three \acp{PW}
(%
  steering angles:
  $\vartheta_{0} = \SI{-20}{\degree}$,
  $\vartheta_{1} = \SI{0}{\degree}$,
  $\vartheta_{2} = \SI{10}{\degree}$%
) for
offline processing.
The excitation voltage was
a single cycle at
\SI{4}{\mega\hertz}.

\subsection{Image Formation}
\input{methods/methods_image_formation.tex}

\subsection{Sampling Grids}
\input{methods/methods_sampling.tex}

\subsection{Image Post-Processing}
\label{subsec:methods_post_processing}
\input{methods/methods_post_processing.tex}


%% file: methods/methods_image_formation.tex
The \ac{DAS} algorithm was executed in
the Fourier domain
\cite{%
  proc:SchiffnerIUS2021,%
  proc:SchiffnerIUS2016a%
} and used
the frequencies between
$f_{\text{lb}} = \SI{2.25}{\mega\hertz}$ and
$f_{\text{ub}} = \SI{6.75}{\mega\hertz}$.
The \ac{FOV} was
an axis-aligned square with
an edge length of
$\SI{39}{\milli\meter}$ and
laterally centered in front of
the linear array with
an axial shift of
$\SI{5}{\milli\meter}$.
The apodization weights derived from
a Tukey window with
a cosine fraction of
\SI{20}{\percent}.
The wavenumber bounds in
the recoverable passbands
\eqref{eqn:theory_spectral_properties_support_pw_single} were
$k_{\text{lb}} = 2 \pi f_{\text{lb}} / c \approx \SI{9187.4}{\radian\per\meter}$ and
$k_{\text{ub}} = 2 \pi f_{\text{ub}} / c \approx \SI{27562.3}{\radian\per\meter}$.
%
The $F$-number of
$F = 1$ limited
the angular aperture and induced
the lower and
upper bounds on
the receive angle
$\varphi_{\text{ub}} = - \varphi_{\text{lb}} \approx \SI{26.6}{\degree}$.
%
%
%
%
The bounds on
the angular spatial frequencies in
the image spectrum
\eqref{eqn:theory_spectral_properties_support_pw_sequence_bounds} were
$\hat{k}_{x, \text{lb}} \approx \SI{-21753.1}{\radian\per\meter}$,
$\hat{k}_{x, \text{ub}} \approx \SI{17112.4}{\radian\per\meter}$,
$\hat{k}_{z, \text{lb}} \approx \SI{16850.9}{\radian\per\meter}$, and
$\hat{k}_{z, \text{ub}} \approx \SI{55124.7}{\radian\per\meter}$.

%% file: methods/methods_sampling.tex
The proposed rhombic grid used
the length
$\Delta u \approx \SI{38273.8}{\radian\per\meter}$ in
the Fourier domain.
The number of
image voxels induced by
the resulting grid vectors
\eqref{eqn:theory_sampling_rhombic_vectors}, which had
the length
$\Delta r \approx \SI{189.6}{\micro\meter}$, amounted to
\num{\NVoxelsRhombic}.
The first orthogonal grid oversampled
the image.
The number of
image voxels and
their spacings amounted to
$\num{512} \times \num{512} = \num{\NVoxelsOrthogonalUsual}$ and
$\Delta x = \Delta z = p / 4 = \SI{76.2}{\micro\meter}$,
respectively.
The second orthogonal grid used
the optimal spacings
\eqref{eqn:theory_sampling_orthogonal_spacing}, which amounted to
$\Delta x \approx \SI{161.7}{\micro\meter}$ and
$\Delta z \approx \SI{164.2}{\micro\meter}$.
The number of
image voxels was
$\num{241} \times \num{237} = \num{\NVoxelsOrthogonalOptimal}$.

%% file: methods/methods_post_processing.tex
All images, after
the coherent compounding, were interpolated to
the first orthogonal grid.
This interpolation simplified
comparisons by
the mean \ac{SSIM} indices
\cite{article:WangISPM2009} and
the relative \acp{RMSE}.
The author maintains
a public version of
the \name{Matlab}%
\footnote{%
  The MathWorks, Inc., Natick, MA, USA
  \label{ftnote:manufacturer_mathworks}
} source code
\cite{software:RhombicGrids} to support
the reproduction of
the presented results and facilitate
further research.
%

%% file: results/results.tex
%
\begin{figure}[t!]
 \centering%
  \input{results/figures/latex/results_experiments_cirs_040.tex}
 \caption{}
 \label{fig:V}
\end{figure}
C
{
 Results for
 the multi-tissue phantom.
 The images show
 the spectra (left column) and
 the absolute voxel values (right column) for
 three steered \acp{PW} and
 \subref{fig:results_experiments_cirs_040_orthogonal_dense_spectrum},
 \subref{fig:results_experiments_cirs_040_orthogonal_dense_image}
 the usual orthogonal grid,
 \subref{fig:results_experiments_cirs_040_orthogonal_optimal_spectrum},
 \subref{fig:results_experiments_cirs_040_orthogonal_optimal_image}
 the optimal orthogonal grid with
 the spacings
 \eqref{eqn:theory_sampling_orthogonal_spacing}, and
 \subref{fig:results_experiments_cirs_040_rhombic_spectrum},
 \subref{fig:results_experiments_cirs_040_rhombic_image}
 the proposed rhombic grid with
 the vectors
 \eqref{eqn:theory_sampling_rhombic_vectors}.
 Both images in
 \subref{fig:results_experiments_cirs_040_orthogonal_optimal_image} and
 \subref{fig:results_experiments_cirs_040_rhombic_image} were interpolated to
 the usual orthogonal grid in
 \subref{fig:results_experiments_cirs_040_orthogonal_dense_image} to simplify
 comparisons.
 The axes in
 both columns are
 identical.
 The angular spatial frequencies were normalized by
 $8 \pi / p$.
}%
{exp_val_cirs_040_images}

The image spectra in
\cref{%
  fig:results_experiments_cirs_040_orthogonal_dense_spectrum,%
  fig:results_experiments_cirs_040_orthogonal_optimal_spectrum,%
  fig:results_experiments_cirs_040_rhombic_spectrum%
} strongly resembled
the theoretical predictions.
The spectrum of
the oversampled image in
\cref{fig:results_experiments_cirs_040_orthogonal_dense_spectrum} equaled
the example in
\cref{fig:theory_passbands_pw_broadband_sequence} except for
artifacts.
These artifacts showed outside
the total recoverable passband
\eqref{eqn:theory_spectral_properties_support_pw_sequence} and stemmed from
the \ac{FOV}, whose boundedness limited
the resolution of
the spectral analysis and caused
spectral leakage.
The image spectra in
\cref{%
  fig:results_experiments_cirs_040_orthogonal_optimal_spectrum,%
  fig:results_experiments_cirs_040_rhombic_spectrum%
} showed
the periodic superposition that was predicted in
\cref{fig:theory_sampling_pw}.
The absence of
overlaps in
these spectra enabled
the recovery of
the spectrum in
\cref{fig:results_experiments_cirs_040_orthogonal_dense_spectrum} and, thus,
the corresponding image in
\cref{fig:results_experiments_cirs_040_orthogonal_dense_image}.
The images in
\cref{%
  fig:results_experiments_cirs_040_orthogonal_optimal_image,%
  fig:results_experiments_cirs_040_rhombic_image%
}, for
this reason, were
visually identical to
the image in
\cref{fig:results_experiments_cirs_040_orthogonal_dense_image}.

\begin{table}[tb!]
 \centering
 \caption{%
  Mean \acf{SSIM} indices and
  relative \acfp{RMSE} achieved by
  the optimal orthogonal grid and
  the proposed rhombic grid.
 }
 \label{tab:conclusion}
 \small
 \begin{tabular}{%
  @{}%
  l
  S[table-format=5.0,table-number-alignment = right,table-auto-round]
  S[table-format=2.1,table-number-alignment = right,table-auto-round]
  S[table-format=1.1,table-number-alignment = right,table-auto-round]
  @{}%
 }
 \toprule
  \multicolumn{1}{@{}H}{Grid type} &
  \multicolumn{1}{H}{Voxel number} & 
  \multicolumn{1}{H@{}}{Mean \acs{SSIM} (\si{\percent})} &
  \multicolumn{1}{H}{Rel. \acs{RMSE} (\si{\percent})}\\
  \cmidrule(r){1-1}\cmidrule(lr){2-2}\cmidrule(lr){3-3}\cmidrule(l){4-4}
 \addlinespace
  Usual orthogonal & \NVoxelsOrthogonalUsual & 100.0 & 0.0\\
  Optimal orthogonal & \NVoxelsOrthogonalOptimal & \MeanSSIMIndexOptimal & \relRMSEOptimal\\
  Proposed & \NVoxelsRhombic & \MeanSSIMIndexRhombic & \relRMSERhombic\\  
 \addlinespace
 \bottomrule
 \end{tabular}
\end{table}

Mean \ac{SSIM} indices above
\SI{\MeanSSIMIndexRhombic}{\percent} and
relative \acp{RMSE} below
\SI{\relRMSERhombic}{\percent} confirmed
the visual equivalence of
all images in
\cref{%
  fig:results_experiments_cirs_040_orthogonal_dense_image,%
  fig:results_experiments_cirs_040_orthogonal_optimal_image,%
  fig:results_experiments_cirs_040_rhombic_image%
}, as summarized in
\cref{tab:conclusion}.
The optimal orthogonal grid achieved
slightly better mean \ac{SSIM} indices and
relative \acp{RMSE} than
the proposed rhombic grid.
This difference was probably caused by
numerical interpolation errors.
The proposed rhombic grid, however, reduced
the number of
image voxels by
\SI{\NVoxelsRelDiffUsual}{\percent} and
\SI{\NVoxelsRelDiffOptimal}{\percent} in comparison to
the usual and
optimal orthogonal grids,
respectively.
Less image voxels usually imply
lower computational costs and
lower memory consumption if
the costs of
the interpolation are
negligible.

%% file: results/figures/latex/results_experiments_cirs_040.tex
%
\begin{tikzpicture}
[
  font = \footnotesize,
  arrow_indicator/.style = { white, thick, ->, >=Stealth },
  title/.style = { inner sep = 0, outer sep = 0, anchor = base, font = \bfseries },
  f_number/.style = { white, anchor = south west },
  no_f_number/.style = { draw = none, fill = RUBGRAY_RGB, rounded corners = 0, minimum width = 0.5em },
  fixed_f_number/.style = { draw = none, fill = RUBGREEN_RGB, rounded corners = 0, minimum width = 0.5em },
  proposed_f_number/.style = { draw = none, fill = RUBBLUE_RGB, rounded corners = 0, minimum width = 0.5em },
  legend/.style = { draw = black!50, fill = black!5, rounded corners = 5, thick, outer sep = 0 }%
]

\pgfmathsetmacro{\AxisLengthX}{3.0}   
\pgfmathsetmacro{\AxisLengthY}{3.0}   

\pgfmathsetmacro{\AxisMetricsLengthX}{3.4}   
\pgfmathsetmacro{\AxisMetricsLengthZNorm}{0.4}

\pgfmathsetmacro{\AxisDistanceX}{1}  
\pgfmathsetmacro{\AxisDistanceY}{1.1}  
\pgfmathsetmacro{\AxisDistanceYColorbar}{1}  

\pgfmathsetmacro{\ColorbarLengthY}{0.2}     
\pgfmathsetmacro{\ColorbarDistanceY}{1.1}   

\pgfmathsetmacro{\BGMarginX}{0.2}
\pgfmathsetmacro{\BGMarginY}{0.2}
\pgfmathsetmacro{\BGTitleY}{1}
\pgfmathsetmacro{\BGTitleX}{1.5}

\pgfmathsetmacro{\ImageLimitsKxLB}{-0.5} 
\pgfmathsetmacro{\ImageLimitsKxUB}{0.5}  
\pgfmathsetmacro{\ImageLimitsKzLB}{0}    
\pgfmathsetmacro{\ImageLimitsKzUB}{1}    

\pgfmathsetmacro{\AxisLimitsKxLB}{-0.5}  
\pgfmathsetmacro{\AxisLimitsKxUB}{0.5}   
\pgfmathsetmacro{\AxisLimitsKzLB}{0}     
\pgfmathsetmacro{\AxisLimitsKzUB}{1}    

\pgfmathsetmacro{\ImageLimitsXLBCIRS}{-19.5072} 
\pgfmathsetmacro{\ImageLimitsXUBCIRS}{19.5072}  
\pgfmathsetmacro{\ImageLimitsZLBCIRS}{4.8768}   
\pgfmathsetmacro{\ImageLimitsZUBCIRS}{43.8912}  

\pgfmathsetmacro{\AxisLimitsXLBCIRS}{-19.25}  
\pgfmathsetmacro{\AxisLimitsXUBCIRS}{19.25}   
\pgfmathsetmacro{\AxisLimitsZLBCIRS}{5.4}     
\pgfmathsetmacro{\AxisLimitsZUBCIRS}{43.9}    

\pgfmathsetmacro{\MagCtrX}{1.92625}         
\pgfmathsetmacro{\MagCtrZ}{25.9644}         
\pgfmathsetmacro{\MagLengthX}{3}            
\pgfmathsetmacro{\MagLengthZ}{3}            
         
\pgfmathsetmacro{\MagImageSizeRel}{0.3}     
\pgfmathsetmacro{\MagImageShiftX}{0.01}     
\pgfmathsetmacro{\MagImageShiftZ}{0.01}     

\pgfmathsetmacro{\MagScaleLengthX}{1}       
\pgfmathsetmacro{\MagScaleLengthZ}{2}       
\pgfmathsetmacro{\MagScalePosRelZ}{0.35}    

\pgfmathsetmacro{\AxisLabelXShift}{-0.2}  
\pgfmathsetmacro{\AxisLabelZShift}{-0.17}  

\pgfmathsetmacro{\ArrowLength}{5}
\pgfmathsetmacro{\ArrowAngle}{125}

\pgfmathsetmacro{\PositionArrowX}{-5}
\pgfmathsetmacro{\PositionArrowZ}{22}

\pgfmathsetmacro{\MagLimitsXLB}{ \MagCtrX - 0.5 * \MagLengthX } 
\pgfmathsetmacro{\MagLimitsXUB}{ \MagCtrX + 0.5 * \MagLengthX } 
\pgfmathsetmacro{\MagLimitsZLB}{ \MagCtrZ - 0.5 * \MagLengthZ } 
\pgfmathsetmacro{\MagLimitsZUB}{ \MagCtrZ + 0.5 * \MagLengthZ } 
\pgfmathsetmacro{\MagImageSizeX}{ \MagImageSizeRel * \AxisLengthX }
\pgfmathsetmacro{\MagImageSizeZ}{ \MagImageSizeRel * \AxisLengthY }
\pgfmathsetmacro{\MagImagePosX}{ \MagImageShiftX * \AxisLengthX }
\pgfmathsetmacro{\MagImagePosZ}{ \MagImageShiftZ * \AxisLengthY }
\pgfmathsetmacro{\MagScaleXLB}{ \MagCtrX - 0.5 * \MagScaleLengthX } 
\pgfmathsetmacro{\MagScaleXUB}{ \MagCtrX + 0.5 * \MagScaleLengthX } 
\pgfmathsetmacro{\MagScaleZ}{ \MagCtrZ + \MagScalePosRelZ * 0.5 * \MagLengthZ }
\pgfmathsetmacro{\MagScaleLengthZImage}{ \MagScaleLengthZ * ( \MagLimitsZUB - \MagLimitsZLB ) / ( 10 * \MagImageSizeRel * \MagLengthZ ) }
\pgfmathsetmacro{\MagScaleZLB}{ \MagScaleZ - 0.5 * \MagScaleLengthZImage }
\pgfmathsetmacro{\MagScaleZUB}{ \MagScaleZ + 0.5 * \MagScaleLengthZImage }

\pgfmathsetmacro{\AxisMetricsLengthZ}{ \AxisMetricsLengthZNorm * \AxisLengthY }   

\pgfmathsetmacro{\ColorbarLengthX}{ 2 * \AxisLengthX + \AxisDistanceX }
\pgfmathsetmacro{\ColorbarDistanceXRel}{ ( \AxisLengthX + \AxisDistanceX ) / \AxisLengthX }
\pgfmathsetmacro{\ColorbarDistanceYRel}{ \ColorbarDistanceY / \AxisLengthY }

\pgfmathsetmacro{\LegendLengthX}{ \ColorbarLengthX }

\pgfmathsetmacro{\AxisPosXZero}{ 0 }
\pgfmathsetmacro{\AxisPosXOne}{ \AxisLengthX + \AxisDistanceX }

\pgfmathsetmacro{\AxisPosYZero}{ 0 }
\pgfmathsetmacro{\AxisPosYOne}{ - \AxisDistanceY - \AxisLengthY }
\pgfmathsetmacro{\AxisPosYTwo}{ 2 * \AxisPosYOne }

\pgfplotsset{%
  b_mode_spectrum/.style = {%
    width = \AxisLengthX cm, height = \AxisLengthY cm,
    scale only axis = true,
    y dir = reverse,
    axis on top,
    axis x line = bottom,
    axis y line = left,
    axis line style = { - },
    every axis title/.append style = { title, at = { ( 0, 1 ) } },
    xmin = \AxisLimitsKxLB, xmax = \AxisLimitsKxUB,
    ymin = \AxisLimitsKzLB, ymax = \AxisLimitsKzUB,
    xtick = { -0.5, -0.25, 0, 0.25, 0.5 },
    xticklabel style = { font = \scriptsize },
    xticklabels = { $-\frac{1}{2}$, $-\frac{1}{4}$, 0, $\frac{1}{4}$, $\frac{1}{2}$ },
    ytick = { 0, 0.25, 0.5, 0.75, 1 },
    yticklabel style = { font = \scriptsize },
    yticklabels = { 0, $\frac{1}{4}$, $\frac{1}{2}$, $\frac{3}{4}$, 1 },
    x label style = { at = { (axis description cs:0.5,\AxisLabelXShift) }, anchor = north, outer sep = 0, inner sep = 0 },
    y label style = { at = { (axis description cs:\AxisLabelZShift,0.5) }, anchor = south, outer sep = 0, inner sep = 0 },
  }
}

\pgfplotsset{%
  b_mode_image/.style = {%
    width = \AxisLengthX cm, height = \AxisLengthY cm,
    scale only axis = true,
    y dir = reverse,
    axis on top,
    axis x line = bottom,
    axis y line = left,
    axis line style = { - },
    every axis title/.append style = { title, at = { ( 0, 1 ) } },
    xtick distance = 5,
    xticklabel style = { font = \scriptsize },
    ytick distance = 5,
    yticklabel style = { font = \scriptsize },
    x label style = { at = { (axis description cs:0.5,\AxisLabelXShift) }, anchor = north, outer sep = 0, inner sep = 0 },
    y label style = { at = { (axis description cs:\AxisLabelZShift-0.01,0.5) }, anchor = south, outer sep = 0, inner sep = 0 },
  }
}

\pgfplotsset{%
  cirs_040_phantom/.style = {%
    b_mode_image,
    xmin = \AxisLimitsXLBCIRS, xmax = \AxisLimitsXUBCIRS,
    ymin = \AxisLimitsZLBCIRS, ymax = \AxisLimitsZUBCIRS,
  }
}

\pgfplotsset{%
  colorbar_common/.style = {%
    point meta min = #1, point meta max = 0,
    colormap/blackwhite,
    colorbar horizontal,
    colorbar style = {
      at = { ( -\ColorbarDistanceXRel, - \ColorbarDistanceYRel ) },
      anchor = north west,
      xtick distance = 10,
      xticklabel style = { font = \tiny },
      xticklabel = { \pgfmathprintnumber{\tick} },
      width = \ColorbarLengthX cm, height = \ColorbarLengthY cm,
      y label style = { at = { (axis description cs:0,0.5) }, anchor = east, rotate = -90, font = \tiny },
      ylabel = { \si{\deci\bel} },
    },
  }
}

\fill [ black, rounded corners = 5, opacity = 0.1 ] ( \AxisPosXZero - \BGMarginX, \AxisPosYZero + \AxisLengthY + \BGMarginY + \BGTitleY ) rectangle ( \AxisPosXZero + \AxisLengthX + \BGMarginX, \AxisPosYTwo - \BGMarginY );
\node [ title ] at ( \AxisPosXZero + 0.5 * \AxisLengthX, \AxisPosYZero + \AxisLengthY + \BGMarginY + 0.6 * \BGTitleY ) { Spectrum };
\draw ( \AxisPosXZero, \AxisPosYZero + \AxisLengthY + \BGMarginY + 0.45 * \BGTitleY ) -- ++ ( \AxisLengthX, 0 );

\fill [ black, rounded corners = 5, opacity = 0.1 ] ( \AxisPosXOne - \BGMarginX, \AxisPosYZero + \AxisLengthY + \BGMarginY + \BGTitleY ) rectangle ( \AxisPosXOne + \AxisLengthX + \BGMarginX, \AxisPosYTwo - \BGMarginY );
\node [ title ] at ( \AxisPosXOne + 0.5 * \AxisLengthX, \AxisPosYZero + \AxisLengthY + \BGMarginY + 0.6 * \BGTitleY ) { Image };
\draw ( \AxisPosXOne, \AxisPosYZero + \AxisLengthY + \BGMarginY + 0.45 * \BGTitleY ) -- ++ ( \AxisLengthX, 0 );

\fill [ black, rounded corners = 5, opacity = 0.1 ] ( \AxisPosXZero - \BGMarginX - \BGTitleX, \AxisPosYZero + \AxisLengthY + \BGMarginY ) rectangle ( \AxisPosXOne + \AxisLengthX + \BGMarginX, \AxisPosYOne - \BGMarginY );
\node [ title, rotate = 90, anchor = north, font = \scriptsize ] at ( \AxisPosXZero - \BGMarginX - 0.95 * \BGTitleX, \AxisPosYOne + \AxisLengthY + 0.5 * \AxisDistanceY ) { Orthogonal grids };
\draw ( \AxisPosXZero - \BGMarginX - 0.75 * \BGTitleX, \AxisPosYOne ) -- ++ ( 0, 2 * \AxisLengthY + \AxisDistanceY );

\node [ title, rotate = 90, anchor = base ] at ( \AxisPosXZero - \BGMarginX - 0.525 * \BGTitleX, \AxisPosYZero + 0.5 * \AxisLengthY ) { Usual };
\draw ( \AxisPosXZero - \BGMarginX - 0.45 * \BGTitleX, \AxisPosYZero ) -- ++ ( 0, \AxisLengthY );

\node [ title, rotate = 90, anchor = base ] at ( \AxisPosXZero - \BGMarginX - 0.525 * \BGTitleX, \AxisPosYOne + 0.5 * \AxisLengthY ) { Optimal };
\draw ( \AxisPosXZero - \BGMarginX - 0.45 * \BGTitleX, \AxisPosYOne ) -- ++ ( 0, \AxisLengthY );

\fill [ black, rounded corners = 5, opacity = 0.1 ] ( \AxisPosXZero - \BGMarginX - \BGTitleX, \AxisPosYTwo - \BGMarginY ) rectangle ( \AxisPosXOne + \AxisLengthX + \BGMarginX, \AxisPosYTwo + \AxisLengthY + \BGMarginY );
\node [ title, rotate = 90, anchor = center ] at ( \AxisPosXZero - \BGMarginX - 0.75 * \BGTitleX, \AxisPosYTwo + 0.5 * \AxisLengthY ) { Proposed }; 
\draw ( \AxisPosXZero - \BGMarginX - 0.45 * \BGTitleX, \AxisPosYTwo ) -- ++ ( 0, \AxisLengthY );


\begin{axis}
[%
  b_mode_spectrum,
  at = { ( \AxisPosXZero cm, \AxisPosYZero cm ) },
  title = { \phantomsubcaption\label{fig:results_experiments_cirs_040_orthogonal_dense_spectrum} (a) },
]

  \addplot [ forget plot ] graphics [ xmin = \ImageLimitsKxLB, xmax = \ImageLimitsKxUB, ymin = \ImageLimitsKzLB, ymax = \ImageLimitsKzUB ]
    { 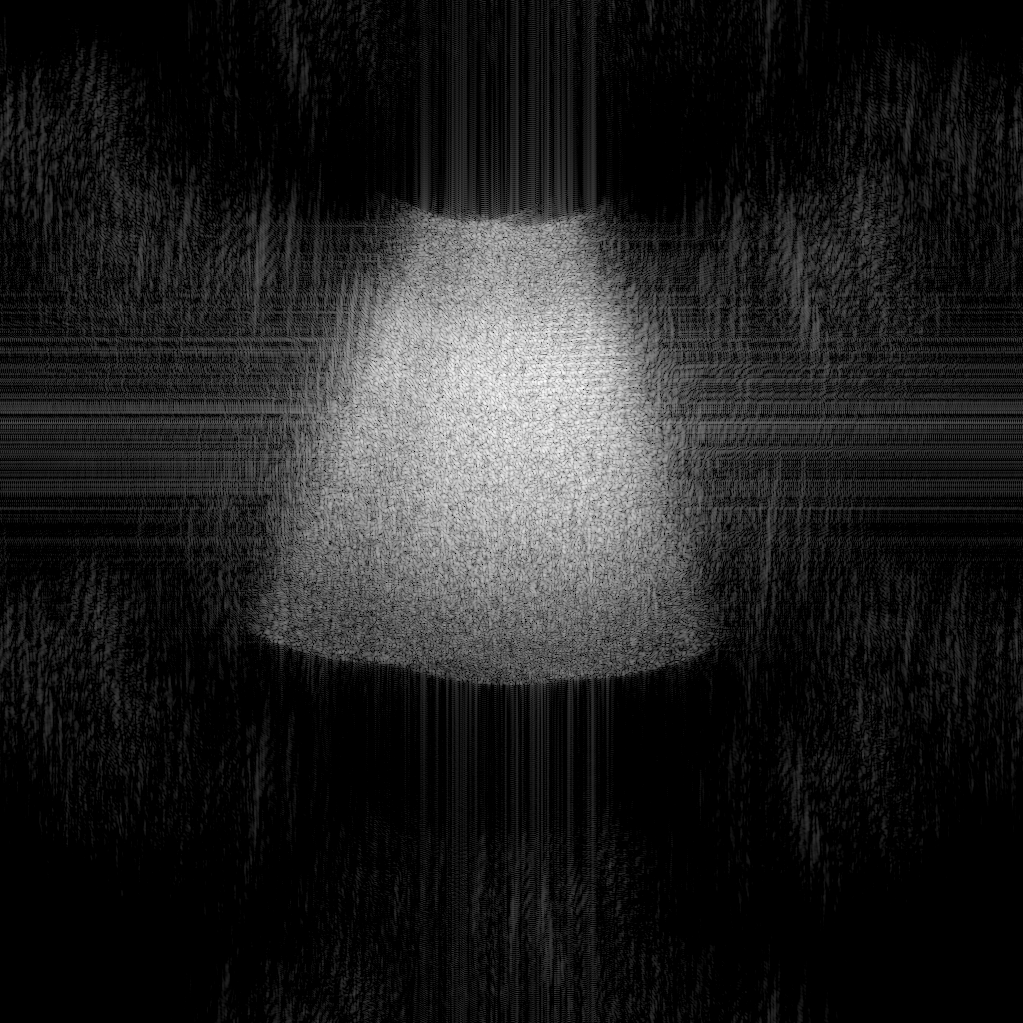 };

%

\end{axis}

\begin{axis}
[%
  cirs_040_phantom,
  at = { ( \AxisPosXOne cm, \AxisPosYZero cm ) },
  title = { \phantomsubcaption\label{fig:results_experiments_cirs_040_orthogonal_dense_image} (b) },
]

  \addplot [ forget plot ] graphics [ xmin = \ImageLimitsXLBCIRS, xmax = \ImageLimitsXUBCIRS, ymin = \ImageLimitsZLBCIRS, ymax = \ImageLimitsZUBCIRS ]
    { 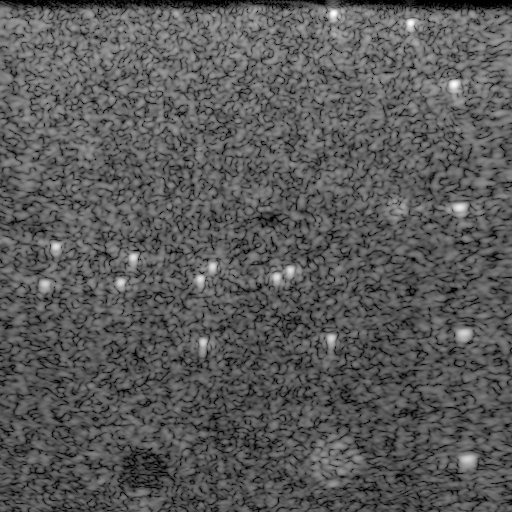 };

%

\end{axis}

\begin{axis}
[%
  b_mode_spectrum,
  at = { ( \AxisPosXZero cm, \AxisPosYOne cm ) },
  title = { \phantomsubcaption\label{fig:results_experiments_cirs_040_orthogonal_optimal_spectrum} (c) },
]

  \addplot [ forget plot ] graphics [ xmin = \ImageLimitsKxLB, xmax = \ImageLimitsKxUB, ymin = \ImageLimitsKzLB, ymax = \ImageLimitsKzUB ]
    { 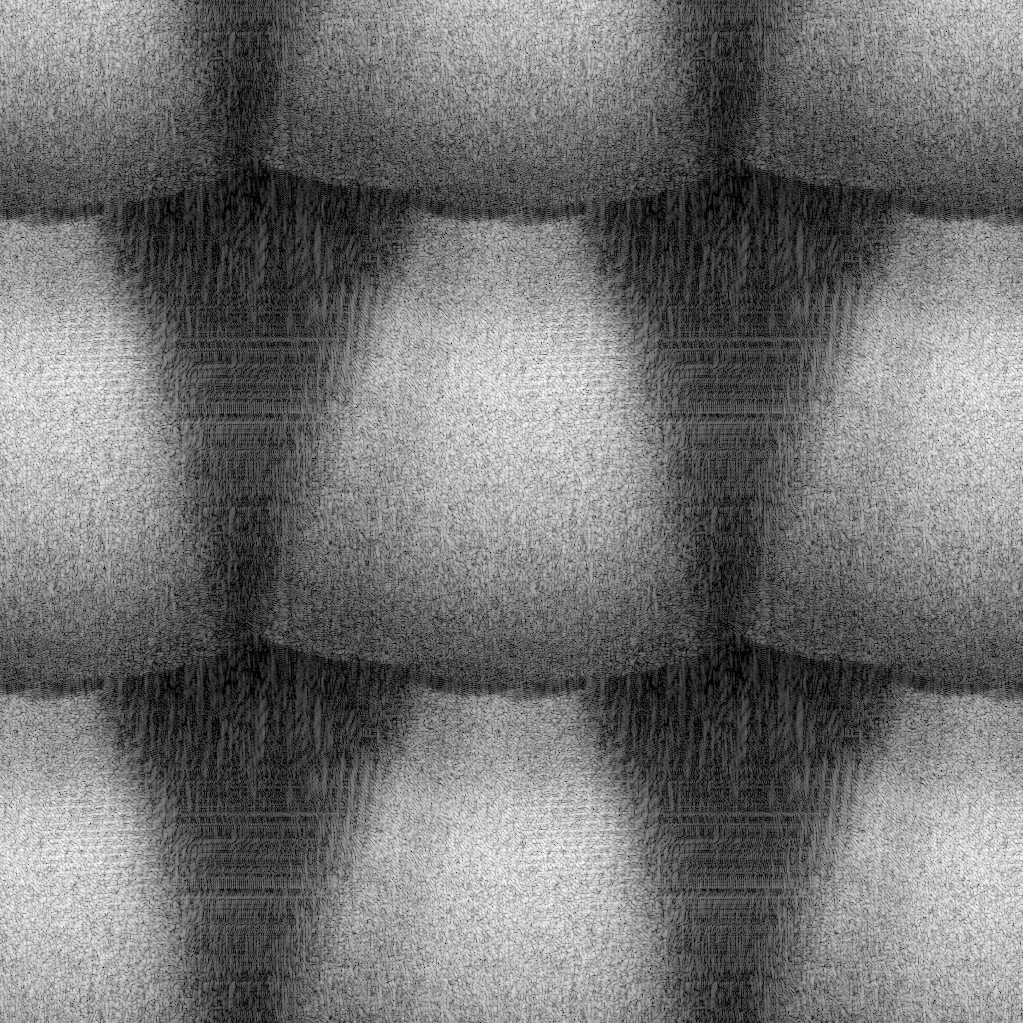 };

%

\end{axis}

\begin{axis}
[%
  cirs_040_phantom,
  at = { ( \AxisPosXOne cm, \AxisPosYOne cm ) },
  xticklabel = { \pgfmathprintnumber{\tick} },
  yticklabel = { \pgfmathprintnumber{\tick} },
  title = { \phantomsubcaption\label{fig:results_experiments_cirs_040_orthogonal_optimal_image} (d) },
]

  \addplot [ forget plot ] graphics [ xmin = \ImageLimitsXLBCIRS, xmax = \ImageLimitsXUBCIRS, ymin = \ImageLimitsZLBCIRS, ymax = \ImageLimitsZUBCIRS ]
    { 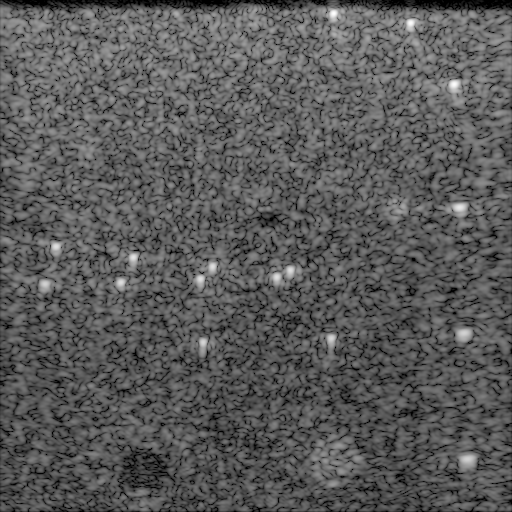 };

%

\end{axis}

\begin{axis}
[%
  b_mode_spectrum,
  at = { ( \AxisPosXZero cm, \AxisPosYTwo cm ) },
  xlabel = { Norm. lateral frequency (1) },
  ylabel = { Norm. axial frequency (1) },
  title = { \phantomsubcaption\label{fig:results_experiments_cirs_040_rhombic_spectrum} (e) },
]

  \addplot [ forget plot ] graphics [ xmin = \ImageLimitsKxLB, xmax = \ImageLimitsKxUB, ymin = \ImageLimitsKzLB, ymax = \ImageLimitsKzUB ]
    { 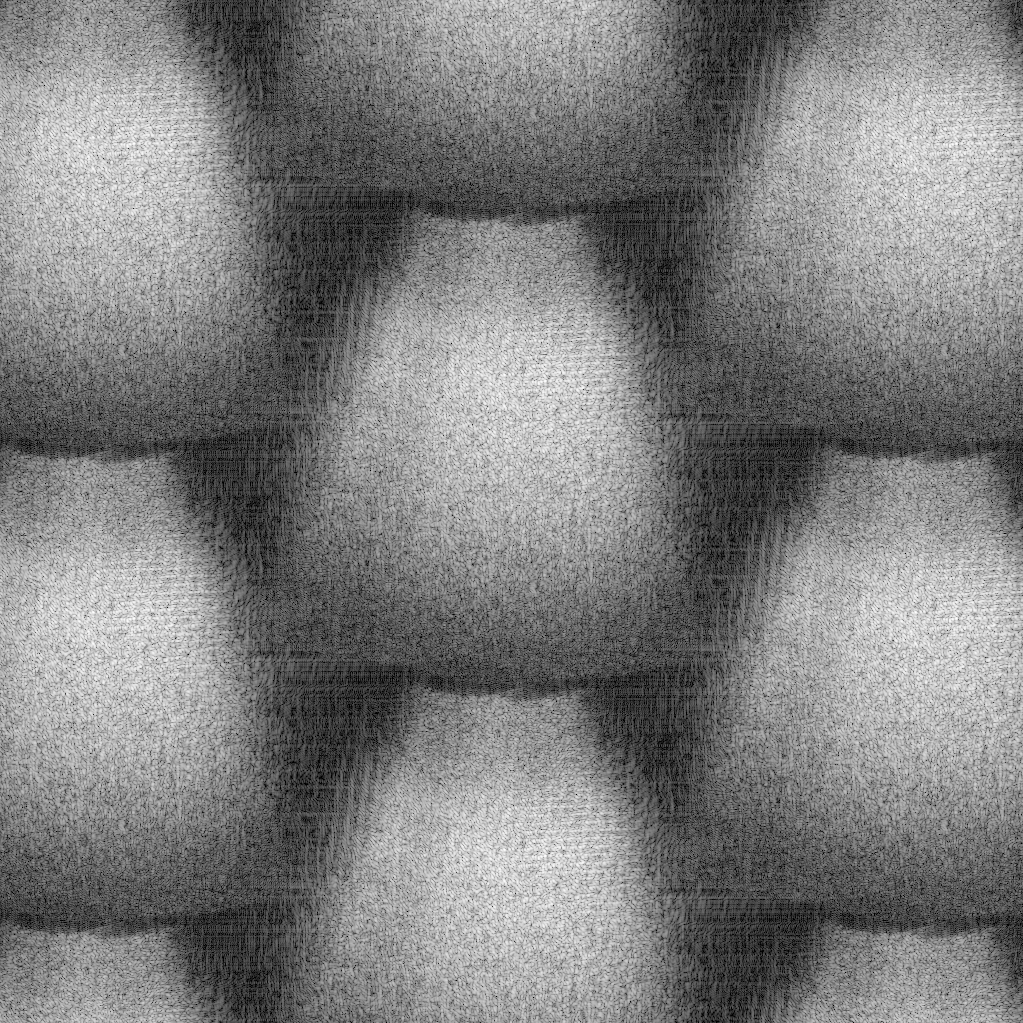 };

%

\end{axis}

\begin{axis}
[%
  cirs_040_phantom,
  at = { ( \AxisPosXOne cm, \AxisPosYTwo cm ) },
  xticklabel = { \pgfmathprintnumber{\tick} },
  yticklabel = { \pgfmathprintnumber{\tick} },
  xlabel = { Lateral position (\si{\milli\meter}) },
  ylabel = { Axial position (\si{\milli\meter}) },
  colorbar_common = { -60 },
  title = { \phantomsubcaption\label{fig:results_experiments_cirs_040_rhombic_image} (f) },
]

  \addplot [ forget plot ] graphics [ xmin = \ImageLimitsXLBCIRS, xmax = \ImageLimitsXUBCIRS, ymin = \ImageLimitsZLBCIRS, ymax = \ImageLimitsZUBCIRS ]
    { 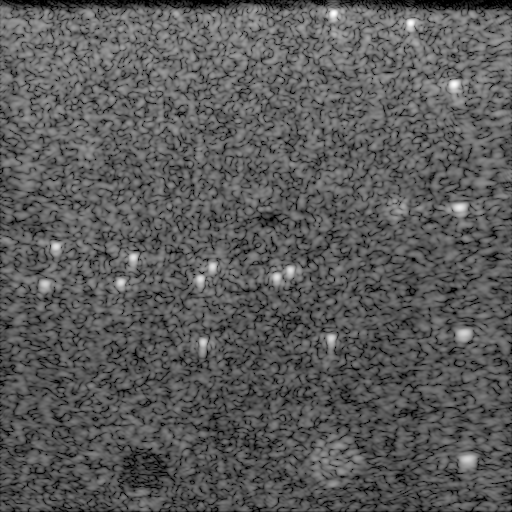 };

%

\end{axis}


\pgfmathsetmacro{\C}{1538.75}  
\pgfmathsetmacro{\P}{304.8}    

\pgfmathsetmacro{\fLB}{2.250}  
\pgfmathsetmacro{\fUB}{6.75}   

\def\configListPWs{-20/RUBGRAY_RGB,0/RUBBLUE_RGB,10/yellow}

\pgfmathsetmacro{\fnumber}{1}

\pgfmathsetmacro{\DeltaX}{ \P / 4 }  

\pgfmathsetmacro{\kLB}{ 2 * pi * \fLB / \C }  
\pgfmathsetmacro{\kUB}{ 2 * pi * \fUB / \C }  

\pgfmathsetmacro{\kLBNorm}{ \fLB * \DeltaX / \C }  
\pgfmathsetmacro{\kUBNorm}{ \fUB * \DeltaX / \C }  

\pgfmathsetmacro{\phiLB}{ - atan( 1 / ( 2 * \fnumber ) ) }
\pgfmathsetmacro{\phiUB}{ - \phiLB }

\coordinate (center) at ( \AxisPosXZero + 0.5 * \AxisLengthX, \AxisPosYZero + \AxisLengthY );

%
\foreach \theta / \col [ count = \indexN from 0 ] in \configListPWs
{

  \pgfmathsetmacro{\klbact}{ \kLBNorm * \AxisLengthX }
  \pgfmathsetmacro{\kubact}{ \kUBNorm * \AxisLengthX }

  \draw [ \col, name path global=setK-\indexN, dash pattern = on 1pt off 1pt ] ($(-90 + \theta:\klbact) + (-90 + \phiLB:\klbact) + (center)$) arc(-90 + \phiLB:-90 + \phiUB:\klbact)
    -- ($(-90 + \theta:\kubact) + (-90 + \phiUB:\kubact) + (center)$) arc(-90 + \phiUB:-90 + \phiLB:\kubact)
    coordinate [ pos = 0.5 ] (pin_\indexN)
    -- cycle;

}

\node [ white, font = \tiny, align = left, inner sep = 0, outer sep = 0 ] (label_passbands) at ( 0.2 * \AxisLengthX, 0.2 * \AxisLengthY ) { Predicted\\passbands };
\draw [ RUBGRAY_RGB, thin ] (label_passbands.north east) -- (pin_0);
\draw [ RUBBLUE_RGB, thin ] (label_passbands.north east) -- (pin_1);
\draw [ yellow, thin ] (label_passbands.north east) -- (pin_2);

\end{tikzpicture}%

%% file: conclusion/conclusion.tex
The proposed rhombic grid
\eqref{eqn:theory_sampling_rhombic_vectors}, as shown in
\cref{tab:conclusion}, maintained
the image quality but reduced
the number of
image voxels by up to
\SI{\NVoxelsRelDiffUsual}{\percent}.
This reduction translated into
reductions in both
the computational costs and
the memory consumption.
The costs of
the additional interpolation
(see \cref{subsec:methods_post_processing}) become
irrelevant if
the number of
steering angles per
compound image is
large enough.
Details of
the theory, such as
(i)
a rigorous derivation of
the recoverable passbands
\eqref{eqn:theory_spectral_properties_support_pw_single} and
\eqref{eqn:theory_spectral_properties_support_pw_sequence} based on
wave acoustics and
(ii)
the effects of
the bounded \ac{FOV}, were left to
an additional publication.
Future research will optimize
the usage of
nonorthogonal regular grids and adapt
the theory to
incident diverging waves and
volumetric \ac{UI}.
Diverging waves are
superpositions of
steered \acp{PW} so that
the theory is
applicable with
only a few modifications.
The author speculates that
the body-centered cubic grid can outperform
orthogonal grids in
the three-dimensional space.
The required extension of
the theory is
simple.